\documentclass[aps,pra,onecolumn,nofootinbib,11pt,tightenlines]{revtex4-1}
\usepackage{amsmath,amssymb,amstext,braket, graphicx,color,xfrac,accents,tabulary,booktabs,xcolor,cases,caption,subcaption}

\usepackage[unicode=true,bookmarks=true,
bookmarksnumbered=false,bookmarksopen=false,
breaklinks=false,
pdfborder={0 0 1},
backref=false,colorlinks=true]{hyperref}
\hypersetup{pdftitle={Quantum Mereology}, pdfauthor={Sean Carroll and Ashmeet Singh}, citecolor=blue,linkcolor=blue,urlcolor=blue,citecolor=blue}
\usepackage{bookmark}

\begin{document}
\baselineskip=14pt
\hfill CALT-TH-2020-023
\hfill

\vspace{0.5cm}
\thispagestyle{empty}

\title{Quantum Mereology: Factorizing Hilbert Space into \\ Subsystems with Quasi-Classical Dynamics}
\author{Sean M. Carroll}
\email{seancarroll@gmail.com}
\author{Ashmeet Singh}
\email{ashmeet@caltech.edu}
\affiliation{Walter Burke Institute for Theoretical Physics, California Institute of Technology, Pasadena, CA 91125}

\newcommand{\be}{\begin{equation}}
\newcommand{\ee}{\end{equation}}
\newcommand{\bea}{\begin{eqnarray}}
\newcommand{\eea}{\end{eqnarray}}
\newcommand{\hst}{\widetilde{\mathcal{H}}} 
\newcommand{\iso}{\dot{=}}
\newcommand{\tdec}{{\{\theta\}}}
\newcommand{\pdec}{{\{\phi\}}}
\newcommand{\spsi}{\ket{\psi}}
\newcommand{\psimu}{\ket{\psi^{(\mu)}}}
\newcommand{\Cmu}{\left[ C^{(\mu)} \right]}
\newcommand{\Y}{\left[ C \right]}
\newcommand{\Ymean}{\left[ \bar{C} \right]}
\newcommand{\deltaY}{\left[ \Delta C \right]}
\newcommand{\OD}{\left[ O_{D} \right]}
\newcommand{\phiM}{\left[ \Phi \right]}
\newcommand{\Dim}{\textrm{dim\,}}
\newcommand{\Tr}{\textrm{Tr\,}}
\newcommand{\hs}{\mathcal{H}} 
\newcommand{\A}{\hat{A}}
\newcommand{\B}{\hat{B}}
\newcommand{\V}{\hat{V}}
\newcommand{\D}{\hat{D}}
\newcommand{\E}{\hat{E}}
\newcommand{\Os}{\hat{O}_{S}}
\newcommand{\Oa}{\hat{O}_{A}}
\newcommand{\pos}{\hat{Q}}
\newcommand{\mom}{\hat{P}}
\newcommand{\ham}{\hat{H}}
\newcommand{\Tu}{\hat{T}_{u}}
\newcommand{\Tv}{\hat{T}_{v}}
\newcommand{\subA}{\mathcal{A}}
\newcommand{\subB}{\mathcal{B}}
\newcommand{\oprho}{\hat{\rho}}
\newcommand{\rhodot}{\dot{\hat{\rho}}_{A}}
\newcommand{\opsig}{\hat{\sigma}}
\newcommand{\Uop}{\hat{U}}
\newcommand{\intham}{\hat{H}_{\rm{int}}}
\newcommand{\selfham}{\hat{H}_{\rm{self}}}
\newcommand{\Ot}{\mathcal{O}(t)}
\newcommand{\Ottwo}{\mathcal{O}(t^2)}
\newcommand{\Otthree}{\mathcal{O}(t^3)}
\newcommand{\eye}{\hat{\mathbb{I}}}
\newcommand{\basisB}{\mathcal{B}}
\newcommand{\trace}{\mathrm{Tr}}
\newcommand{\Aa}{\hat{A}_{\alpha}}
\newcommand{\Bb}{\hat{B}_{\beta}}
\newcommand{\tp}{\otimes}
\newcommand{\lla}{\left\langle}
\newcommand{\rra}{\right\rangle}
\newcommand{\lcb}{\left[}
\newcommand{\rcb}{\right]}
\newcommand{\Oj}{\hat{O}_{j}}
\newcommand{\opphi}{\hat{\phi}}
\newcommand{\oppi}{\hat{\pi}}
\newcommand{\M}{\hat{M}}

\newcommand{\sean}[1]{\textbf{\color{red}{#1}}}
\newcommand{\ashmeet}[1]{\textbf{\color{blue}{#1}}}
\newcommand{\todo}[1]{\textbf{\color{orange}{#1}}}
\begin{abstract}
We study the question of how to decompose Hilbert space into a preferred tensor-product factorization without any pre-existing structure other than a Hamiltonian operator, in particular the case of a bipartite decomposition into ``system'' and ``environment."
Such a decomposition can be defined by looking for subsystems that exhibit quasi-classical behavior.
The correct decomposition is one in which pointer states of the system are relatively robust against environmental monitoring (their entanglement with the environment does not continually and dramatically increase) and remain localized around approximately-classical trajectories. 
We present an in-principle algorithm for finding such a decomposition by minimizing a combination of entanglement growth and internal spreading of the system.
Both of these properties are related to locality in different ways.
This formalism is relevant to questions in the foundations of quantum mechanics and the emergence of spacetime from quantum entanglement.
\end{abstract}
\maketitle
\newpage
\tableofcontents
\newpage
\section{Introduction}

If someone hands you two qubits $A$ and $B$, there is a well-understood procedure for constructing the quantum description of the composite system constructed from the two of them.
If the individual Hilbert spaces are $\hs_A\simeq \mathbb{C}^2$ and $\hs_B\simeq \mathbb{C}^2$, the composite Hilbert space is given by the tensor product, $\hs \simeq \hs_A\otimes \hs_B \simeq \mathbb{C}^4$, where $\simeq$ represents isomorphism.
The total Hamiltonian is the sum of the two self-Hamiltonians, $\ham_A$ and $\ham_B$, acting on $\hs_A$ and $\hs_B$, respectively, plus an appropriate interaction term, $\intham$, coupling the two factors.

What about the other way around?
If someone hands you a four-dimensional Hilbert space and a Hamiltonian, is there a procedure by which we can factorize the system into the tensor product of two qubits?
In general there will be an infinite number of possible factorizations, each defined by a bijection of the form 
\be
\lambda : \hs \rightarrow \hs_A\otimes \hs_B .
\ee
Unitary transformations $\hat U$ can be used to define different bijections, 
\be
\tilde\lambda = \lambda \circ \hat{U} : \hs \rightarrow \hs_A\otimes \hs_B.
\ee
While some unitaries will simply induce rotations within the factors $\hs_A$ and $\hs_B$, generically the factorization defined by $\tilde\lambda$ will not be equivalent to that defined by $\lambda$.
Is there some notion of the ``right'' factorization for a given physical situation?

In almost all applications, these questions are begged rather than addressed.
When someone hands us two spin-1/2 particles, it seems obvious how to assign Hilbert spaces to each and form the relevant tensor product.
But there are circumstances,when we might know nothing more than the total Hilbert space and the Hamiltonian (and perhaps a specified initial state), and want to use that information to reverse-engineer a sensible notion of what physical system is being described, including what its individual parts are.
This is the subject of ``Quantum Mereology," where ``mereology" is the study of how parts relate to the whole.
It is especially important in the context of finite-dimensional Hilbert spaces, where any Hermitian operator defines an observable, and there is no notion of preferred observables that can be used to define a corresponding factorization.
\\
\\
{This is an important problem for the foundations of quantum mechanics, in particular within an Everettian (Many-Worlds) framework.
The basic postulates of this approach are that the world is fully described by a vector in Hilbert space, evolving according to the Schr\"odinger equation with some given Hamiltonian.
Semi-classical ``worlds'' branch off from each other when quantum systems in superposition become entangled with the environment.
Such a description clearly relies on a factorization of Hilbert space into system and environment, which is usually taken as assumed.
But that factorization is a human convenience, not part of the basic formulation of the theory itself.
We would therefore like to have a set of objective criteria for starting the the fundamental ingredients -- Hilbert space, state, and Hamiltonian -- and deriving everything else, including the appropriate factorization \cite{Carroll:2018rhc}.}

{Another motivation comes from quantum gravity.
The holographic principle \cite{Hooft:1993gx,Susskind:1994vu}, black-hole complementarity \cite{Susskind:1993if}, and the AdS/CFT correspondence \cite{Maldacena:1997re} provide evidence that the ultimate theory of quantum gravity does not arise from a straightforward quantization of general relativity or any other theory with strictly local degrees of freedom.
String theory is an attempt to posit what those degrees of freedom might be, but it might be fruitful to tackle the question from a more general perspective, asking how states in Hilbert space evolving in certain ways can predict emergent behavior describing semiclassical spacetimes obeying Einstein's equation \cite{Banks:2011av,Giddings:2015lla,Cao:2016mst,Cao:2017hrv}.
For such a program, it is crucial to be able to factorize would-be classical states into system and environment, without such a decomposition necessarily being given ahead of time.}

In this paper we seek to address this problem in a systematic way.
Given nothing more than a Hilbert space of some dimensionality, the Hamiltonian, and an initial state, what is the best way to factorize Hilbert space into subsystems?
Since we are not given a preferred factorization to begin with, there is no preferred basis other than the eigenstates of the Hamiltonian.
The Hamiltonian itself is therefore specified by its spectrum (the set of energy eigenvalues), and the initial state by its components in the energy eigenbasis.
Our task is to use this meager data to find the most useful way of decomposing Hilbert space into tensor factors. 

The key here is ``useful,'' and we interpret this as meaning ``allows for a quasi-classical description of the dynamics within the subsystems (or one subsystem coupled to an environment)."
A well-understood feature of conventional quantum dynamics is the selection of pointer states of a system that is being monitored by an environment. 
In general the reduced density matrix of the system can always be diagonalized in some basis, but for systems that can exhibit quasi-classical behavior, the pointer states define a basis in which the system's density matrix will rapidly approach a diagonal form.
These pointer states then obey quasi-classical dynamics.
This implies in particular that a system in a pointer state remains relatively unentangled with the environment, and that we can define pointer observables that approximately obey classical equations of motion. 
This suggests a criterion for determining the proper system/environment factorization: choose the tensor-product decomposition in which the system has a pointer basis that most closely adheres to these properties.
As we will see, generic Hamiltonians will have no such decomposition available, so quasi-classical behavior is non-generic.

In this paper we develop an algorithm for making this criterion precise.
For any given decomposition, we start with an unentangled state, and calculate the growth of entanglement.
Since our interest is in finite-dimensional Hilbert spaces \cite{Bao:2017rnv,Banks2000,Fischler2000}, we use Generalized Pauli Operators (which have their algebraic roots in generalized Clifford algebra) to define conjugate operators $\hat q$ and $\hat p$; in the infinite-dimensional limit, these obey the Heisenberg canonical commutation relations. 
The position operator $\hat q$ is the one that appears in the interaction Hamiltonian.
We can then calculate the rate of spread of the uncertainty in the position variable.
Both the entanglement between system and environment and the spread of the system's position can be characterized by an entropy.
Our criterion is that the correct decomposition minimizes the maximum of these two entropies, for initially localized and unentangled states.

While this question has not frequently been addressed in the literature on quantum foundations and emergence of classicality, 
a few works have highlighted its importance and made attempts to understand it better. Brun and Hartle \cite{brun1999classical} studied 
the emergence of preferred coarse-grained classical variables in a chain of quantum harmonic oscillators. Efforts to address the closely related 
question of identifying classical set of histories (also known as the ``Set Selection" problem) in the Decoherent Histories 
formalism \cite{gell2019alternative,dowker1996consistent,kent1998quantum,riedel2012,Griffiths:1984rx,paz1993environment} have also been undertaken.
 Tegmark \cite{tegmark2015consciousness} has approached the problem from the perspective of information processing ability of subsystems and
 Piazza \cite{Piazza:2005wm} focuses on emergence of spatially local subsystem structure in a field theoretic context. 
Hamiltonian induced factorization of Hilbert space which exhibit $k$-local dynamics has also been studied by Cotler
 \emph{et al} \cite{cotler2019locality}). The idea that tensor product structures and virtual subsystems can be identified with algebras
 of observables was originally introduced by Zanardi \emph{et al} in \cite{Zanardi:2001zz,Zanardi:2004zz} and was further extended in
 Kabernik, Pollack and Singh \cite{Kabernik:2019jko} to induce more general structures in Hilbert space. In a series of papers
 (e.g.\ \cite{2004SHPMP..35...73C,Castagnino:2008zz,10.1007/978-94-007-2404-4_15,2014BrJPh..44..138F}; see also \cite{2005PhRvA..72a2109S})
 Castagnino, Lombardi, and collaborators have developed the self-induced decoherence (SID) program, which conceptualizes decoherence as a
 dynamical process which identifies the classical variables by inspection of the Hamiltonian, without the need to explicitly identify a
 set of environment degrees of freedom. Similar physical motivations but different mathematical methods have led Kofler and 
Brukner \cite{2007PhRvL..99r0403K} to study the emergence of classicality under restriction to coarse-grained measurements.

The paper is organized as follows. Section \ref{sec:qc_factorization} describes the important features of a quasi-classical factorization, settling on two important features: ``robustness," referring to slow growth of entanglement between pointer states and the environment, and ``predictability," meaning that pointer observables approximately obey classical equations with low variance.
We emphasize how these features will not be manifest in any arbitrary factorization and use a bipartite example to demonstrate these characteristics.
We then examine these two features in turn.
Section \ref{sec:robustness} considers robustness, showing that it is non-generic, and investigating what kinds of decompositions will minimize the growth of entanglement.
In section \ref{sec:predictability}, we discuss predictability of classical states under evolution, and outline a method to quantify the spread induced in initially predictable states of the pointer observable.
In section \ref{sec:algorithm}, we will outline an algorithm to sift through different decompositions of Hilbert space, given a Hamiltonian to pick out the one with manifest quasi-classicality. We will define an entropy-based quantity that we call \emph{Schwinger Entropy} whose minimization ensures the existence of low entropy states that are both resistant to entanglement production and have a pointer observable that evolves quasi-classically. In Section \ref{sec:conjugate}, we make contact with the notion of conjugate variables, and describe under what conditions one can interpret them as classical position and momentum. We connect quasi-classicality with features of the Hamiltonian. These include the ``collimation'' of the self-Hamiltonian, needed to ensure that initially peaked states remain relatively peaked, and the pointer observables approximately obeying classical equations of motion.
We close with a worked example and some discussion.

\section{Factorization and Classicality}
\label{sec:qc_factorization}

There is a great deal of freedom in the choice of factorization of Hilbert space corresponding to different subsystems.
In principle any factorization can be used, or none; for purposes of unitary dynamics, one is free to express the quantum state however one chooses.
For purposes of pinpointing quasi-classical behavior, however, choosing the right factorization into system $\mathcal{S}$ and environment $\mathcal{E}$ is crucial.
Similar considerations will apply to further factorization of the system into subsystems.
Let us therefore review what is meant by ``quasi-classical behavior."

Consider a bipartite split of a finite-dimensional Hilbert space $\hs \equiv \left( \subA \otimes \subB \right)_{\tdec}$ into subsystems $\subA$ and $\subB$ in a factorization labeled by $\tdec$ relative to some arbitrary chosen one. 
(In Appendix~\ref{app:factor} we establish some notation and fomulae relevant to factorizations and transformations between them.)
The dimension of $\subA$ is $\Dim \subA = d_{A}$ and $\Dim \subB = d_{B}$, with $\Dim \hs = D = d_{A} d_{B}$. {There is some freedom in the choice of $d_{A}$ and $d_{B}$ as long as they satisfy $d_{A}d_{B} = D$, for a fixed dimension $D$ of the global Hilbert space. We assume that $D$ is non-prime, and hence can be factorized further, as required to obtain subsystem structure. In this work, we focus on the case where $d_{A}$ and $d_{B}$ are a priori fixed, or specified by certain conditions. For instance, in the case where $\subA$ is a quantum system of interest, and $\subB$ a macroscopic environment, we expect $d_{B} \gg d_{A}$.} {It seems likely that our procedure could be extended to fix the best choice of system dimensionality, but we don't explore that here.}

The Hamiltonian $\ham$ in this decomposition can be written as a sum of self terms and an interaction term, following Eq. (\ref{M_self_int}),
\begin{equation}
\label{H_bipartite}
\ham = \ham_{A} \otimes \eye_{B} + \eye_{A} \otimes \ham_{B} + \intham \: .
\end{equation}
We only consider traceless Hamiltonians, so there is no need for  a trace term $h_{0} = \Tr \ham /D $.
Under factorization changes, even though $\Tr \ham$ is preserved, there would be an ambiguity in assigning the trace terms to either of the self-Hamiltonians of $\subA$ or $\subB$. Also, since we are not considering gravity as an external field, subtracting off a constant from $\ham$ is physically trivial. 

The form of the Hamiltonian is dependent on the choice of the decomposition $\tdec$. The interaction term can be expanded in the $SU(d_{A}) \otimes SU(d_{B})$ operator basis as following Eq. (\ref{M_mu_expansion}),
\begin{equation}
\label{Hint_SU}
\intham = \sum_{a = 1}^{d^{2}_{A} - 1} \sum_{b = 1}^{d^{2}_{B} - 1} h_{ab} \left( \hat{\Lambda}^{(A)}_{a} \otimes \hat{\Lambda}^{(B)}_{b} \right) \: .
\end{equation}
One can rewrite $\intham$ in a diagonal form, 
\begin{equation}
\label{Hint_diagonal}
\intham = \sum_{\alpha = 1}^{n_{int}} \lambda_{\alpha}\left( \A_{\alpha} \otimes \B_{\alpha}\right) \: ,
\end{equation}
where $\A_{\alpha}$ and $\B_{\alpha}$ are combinations of the Hermitian generators\footnote{While in a general diagonal decomposition of the interaction Hamiltonian, the operators $\A_\alpha$ and $\B_\alpha$ can be unitary but not necessarily Hermitian, but our form of Eq. (\ref{Hint_diagonal}) is obtained by relabeling/recollecting terms in an expansion with Hermitian terms of Eq. (\ref{Hint_SU}), hence $\A_\alpha$ and $\B_\alpha$ will be Hermitian. This will also help us make easy contact with talking about observables being monitored by subsystems.}
 in Eq. (\ref{Hint_SU}) and the total number of terms will generically be $n_{int} = (d^{2}_{A} - 1)(d^{2}_{B} - 1)$. The coefficients $\lambda_{\alpha}$ characterize the strength of each contribution in the interaction Hamiltonian, which we ensure by absorbing any normalization of operators $\A_{\alpha}$ and $\B_{\alpha}$ in $\lambda_{\alpha}$ such that $|| \A_{\alpha} || = || \B_{\alpha} || = 1$ under a suitable choice of operator norm $|| . ||$. While there appear to be a large number of terms in the expansion in Eq. (\ref{Hint_diagonal}), we will see later how in the preferred, quasi-classical decomposition, most of these terms condense into familiar local operators that serve as pointer observables. 

 A  {quasi-classical} (QC) factorization of $\hs$ that we will denote by $\tdec_{QC}$ can be associated with the following features:
\begin{enumerate}
\item{\textbf{Robustness:} There exist preferred pointer states of the system (and associated pointer observables) that, if initially unentangled with the environment, typically remain unentangled under evolution by $\ham$.}
\item{\textbf{Predictability:} For states with near definite value of the pointer observable, it will serve as a predictable quasi-classical variable, with minimal spreading under Hamiltonian evolution.}
\end{enumerate}
Informally, these two criteria correspond to the conventional notions that ``wave function branchings are rare" and ``expectation values of observables remain peaked around classical trajectories in the appropriate regime."
We can now examine in detail how these features can be characterized quantitatively.

\section{Robustness and Entanglement}
\label{sec:robustness}

It is a feature of the universe (albeit as-yet imperfectly explained) that entropy was low at early times, and has been subsequently increasing \citep{penrose1979singularities,albert}.
In the quantum context, this corresponds to relatively small amounts of initial entanglement between subsystems, and between macroscopic systems and their environment.
Here we are imagining a bipartite split
\begin{equation}
\hs = \mathcal{S} \otimes \mathcal{E}
\end{equation}
into $\mathcal{S}$, which corresponds to ``system'' degrees of freedom we wish to track, and an environment $\mathcal{E}$, which is the part we are not interested in or do not have control over.
In Everettian quantum mechanics \cite{everett1957relative}, this feature underlies the fact that the wave function branches as time moves toward the future, not the past.
Our interest is therefore in initially low-entropy situations, where the system is unentangled with its environment.

With a generic Hamiltonian in a generic factorization, we would expect any initially-unentangled system state to quickly become highly entangled with its environment, on timescales typical of the overall Hamiltonian.
By ``highly entangled'' we mean that the entropy of the system's reduced density matrix would approach $\log(\mathrm{dim}\; \hs_\mathcal{S})$.
In Everettian language, that would correspond to splitting into a number of branches of order $\mathrm{dim}\; \hs_\mathcal{S}$.
This is not what we expect from robust quasi-classical behavior; to a good approximation, Schr\"odinger's cat splits into two branches, not into the exponential of Avogadro's number of branches.

We will therefore ask, given some Hamiltonian $\ham$, how we can factorize $\hs$ into $ \mathcal{S} \otimes \mathcal{E}$ such that the entanglement growth rate of certain initially-unentangled states is minimized.
We will explicitly work to $\mathcal{O}(t^2)$, which we will see is the lowest non-trivial contribution to the entanglement growth. This will help us quantify robustness and quasi-classicality for small times. (Factorizations that are not quasi-classical for small times will not be quasi-classical for later times either.)
 
\subsection{Decoherence Is Non-Generic}
\label{sec:decoherence_feature}
 
It is well-known that wave functions tend to ``collapse'' (or branch) into certain preferred pointer states, depending on what observable is being measured. The decoherence paradigm outlines how in an appropriate factorization, we search for a  {pointer observable} $\Os \in \mathcal{L}(\mathcal{S}) $ such that eigenstates $\{\ket{s_{j}} \: | \: j = 1,2,\cdots,d_{S} \}$ of $\Os$   serve as  {pointer states} \cite{Zurek:1981xq}, which are robust to entanglement production with states of the environment. Thus, there exist special product states $\ket{s_{j}} \otimes \ket{E}$ that do not entangle (or stay approximately unentangled) under the evolution by the total Hamiltonian $\ham$. This feature allows suppression of interference between superpositions of different pointer states, and in the eigenbasis of $\Os$, the reduced density operator for $\mathcal{S}$ given by $\oprho_{S}(t)$ evolves toward a diagonal form, since the conditional environmental states corresponding to different pointer states of the system become  {dynamically} orthogonal $\braket{E(s_j)|E(s_{k})} \to \delta_{jk}$ relatively fast in time. 

In particular in the {Quantum Measurement Limit} (QML) \cite{2007dqct.book.....S}, when the Hamiltonian is dominated by interactions $\intham$ (when the spectral frequencies available in $\intham$ are much larger than those of the self term $\selfham$), the pointer observable satisfies Zurek's commutativity criterion \cite{Zurek:1981xq},
\begin{equation}
\label{Zurek_commutativity_QML}
\lcb \intham , \Os\rcb \approx 0 \: \implies \: \lcb \intham , \Os \otimes \eye_{E} \rcb \approx 0 \: .
\end{equation}  
This is interpreted as saying that the environment $\mathcal{E}$ robustly monitors \cite{joos1985emergence} a certain observable $\Os$ of the system (typically a ``local'' one, such as position) that is compatible with the interaction Hamiltonian $\intham$ and selects this to serve as the pointer observable.\footnote{{There is a related notion of entanglement robustness associated with coherent states (which are eigenstates of the non-hermitian annihilation operator) in the context of
quantum optics (\cite{PhysRevA.65.043605} and references therein). In this paper, we focus on the robustness of pointer states of system observables being monitored by an environment (for a given Hamiltonian) in the context of Zurekian decoherence, and their emergent classicality.}} This commutativity criterion of Eq. (\ref{Zurek_commutativity_QML}) further implies that generically all terms $\Aa$ occurring with $\lambda_{\alpha} \neq 0$ will individually satisfy
\begin{equation}
\lcb \Aa,\Os \rcb \approx 0 \: \: \forall \: \: \alpha \: .
\end{equation}
 The discussion can be extended to the quantum limit of decoherence \cite{2007dqct.book.....S}, where the self term $\selfham$ dominates over $\intham$ and selects eigenstates of the self-Hamiltonian for the system $\ham_{S}$ to be the pointer states. In general, Zurek's ``predictability sieve'' \cite{Zurek:1994zq} sifts through different states in the system's Hilbert space $\mathcal{S}$ to search for states that are robust to entanglement production under evolution by the full Hamiltonian $\ham$. In this paper, we primarily focus on the quantum measurement limit (QML) since a broad class of physical models exhibit this feature where interactions play a dominant and crucial role in the emergence of classicality.
 
Decoherence and the existence of low-entropy states in $\hs$ that do not get entangled under the action of $\ham$ depend sensitively on the Hamiltonian and factorization  $\hs = \subA \otimes \subB$ taking a particular, non-generic form. In the quasi-classical factorization, we will identify subsystem $\subA$ as the ``system" $\mathcal{S}$, and the subsystem $\subB$ as the ``environment" $\mathcal{E}$. In general, as we saw in Eq. (\ref{Hint_SU}) and particularly in the diagonal decomposition Eq. (\ref{Hint_diagonal}), the interaction term has a slew of non-commuting terms $\Aa$ in the summand. Searching for a ``pointer observable" is equivalent to finding an operator compatible with $\intham$, and hence satisfying $\lcb \intham, \hat{O}\rcb \approx 0$. Due to the presence of large number of non-commuting terms in $\intham$, the eigenstates of $\hat{O}$ will be highly entangled and not be low entropy states that can be resilient to entropy production. 

Said differently, the ``pointer observable" $\hat{O}$ will not be of a separable form $\hat{O} \neq \hat{O}_{A} \otimes \hat{O}_{B}$, and only specific factorizations for Hamiltonians can allow decoherence, where many terms of $\intham$ in Eq. (\ref{Hint_diagonal}) conspire together to collect into a  few local and compatible terms allowing for consistent monitoring of the system by the environment.
(To emphasize, by ``decoherence'' here we mean not simply ``becoming entangled with the environment," but the existence of a preferred set of pointer states that define a basis in which the reduced density matrix dynamically diagonalizes.)
As we saw, the existence of initial low entropy states $\oprho(0) = \oprho_{A}(0) \otimes \oprho_{B}(0)$ that are robust under evolution to entanglement production is highly constrained and only in particular cases when many of the $\lambda_{\alpha}$ strengths vanish or terms conspire to condense into a few local terms will they exist to serve as the pointer states for subsystems being robustly monitored by the environment (other subsystems). This can be further understood by considering the constraint counting discussed in Section \ref{subsec:min_entropy} below.

In Appendix~\ref{app:non-generic} we detail this behavior more explicitly.
 
 \subsection{Minimizing Entropy Growth}
\label{subsec:min_entropy}

In an arbitrary decomposition  $\hs \equiv \left( \subA \otimes \subB \right)_{\tdec}$, let us begin with an initial $(t = 0)$ pure state of zero entropy for the factors, which we take to be a product state,
\begin{equation}
\label{initstate}
\oprho(0) \equiv \ket{\psi(0)}\bra{\psi(0)} = \oprho_{A}(0) \otimes \oprho_{B}(0) \equiv \ket{\psi_{A}(0)}\bra{\psi_{A}(0)} \otimes \ket{\psi_{B}(0)}\bra{\psi_{B}(0)} \: .
\end{equation}
At this stage, the decomposition $\tdec$ is completely general and has no notion of preferred observables or classical behavior.  Let us work with a traceless Hamiltonian of Eq. (\ref{H_bipartite}) even though the calculation below holds for $\Tr \ham \neq 0$ since this will only be an overall phase in the unitary evolution of density matrices and hence, cancels out. 
Time evolution of states is implemented using a unitary operator $\hat{U}(t) \equiv \exp{\left( - i \ham t \right)}$, where we are working in units with $\hslash = 1$, and the time evolved state is $\ket{\psi(t)} = \hat{U}(t) \ket{\psi(0)}$. Let us write $\hat{U}(t)$ in a more suggestive form working explicitly to order $\mathcal{O}(t^2)$. 
 
 In Appendix~\ref{sec:low_entropy}, we compute the linear entanglement entropy\footnote{A common entanglement measure used is the von-Neumann entanglement entropy $S_{vN} (\oprho) = - \Tr \left( \oprho \log{\oprho} \right)$ for a given density matrix $\oprho$. However, the presence of the logarithm makes the entropy hard to analytically compute and give expressions for, hence we will focus on its leading order contribution, the Linear Entropy (which is the Tsallis second order entropy measure), $S_{lin}(\oprho) = \left( 1 - \Tr \oprho^{2} \right)$. } $S_{lin}(\oprho_{A}(t)) = \left( 1 - \Tr \oprho^{2}_{A}(t) \right)$ for the reduced density matrix of $\subA$ given by Eq. (\ref{rho_t_2}), which corresponds to starting with an unentangled (and hence, zero entropy) state $\oprho(0)$. 
 Putting these together in Eq. (\ref{Slin_2}), we obtain,
 \begin{equation}
 \label{Slin}
 \begin{split}
 S_{lin}(\oprho_{A}(t))  =  &\  t^2  \sum_{\alpha,\beta} \lambda_{\alpha}\lambda_{\beta} \biggl(  \lla \Aa \A_{\beta} \rra_{0} \lla \B_{\alpha} \Bb \rra_{0} +   \lla \A_{\beta} \Aa \rra_{0} \lla  \Bb \B_{\alpha} \rra_{0}    \\
 &  - \lla \Aa \rra_{0} \lla \A_{\beta} \rra_{0} \left( \lla \{ \B_{\alpha} ,\Bb \}_{+} \rra_{0} -  \lla \B_{\alpha} \rra_{0} \lla \Bb \rra_{0} \right) \\
&  -   \lla \B_{\alpha} \rra_{0} \lla \Bb \rra_{0} \left( \lla \{ \A_{\alpha} ,\A_{\beta} \}_{+} \rra_{0} - \lla \Aa \rra_{0} \lla \A_{\beta} \rra_{0} \right) \biggr) + \Otthree  \:  \: .
 \end{split}
 \end{equation}
 For condensed notation, let us write $ S_{lin}(\oprho_{A}(t)) = \ddot{S}_{lin}(0) \: t^2 + \Otthree$. The quantity $\ddot{S}_{lin}$ will play an important role in quantifying the quasi-classicality of different factorizations of Hilbert space. In particular, for the important case when the interaction Hamiltonian takes the simple form $\intham = \lambda \left(\A \otimes \B\right)$, we notice that the expression for $S_{lin}$ simplifies to,
 \begin{equation}
 \label{Slin_oneterm}
  S_{lin}(\oprho_{A}(t)) \:  = \:  2 \lambda^{2} t^2\left( \lla \A^2 \rra_{0}  - \lla \A\rra^{2}_{0}\right)\left( \lla \B^2 \rra_{0}  - \lla \B\rra^{2}_{0}\right) \: .
 \end{equation}
 
 Let us note a few key features of the entropy growth Eq. (\ref{Slin}). We are working in the context of unentangled (low entropy) states. 
 As we have seen, the entanglement growth rate depends on the interaction strengths $\lambda_{\alpha}$; stronger interactions would entangle subsystems more quickly. One might be temped to conclude that finding a decomposition where the interaction Hamiltonian has the minimum strength would ensure least entanglement production, but we must take note of the important role played by the initial (unentangled) state in determining the rate of entanglement generation\footnote{This is in line with Tegmark's \cite{tegmark2015consciousness} ``Hamiltonian Diagonality Theorem," which proves that the Hamiltonian is maximally separable (with minimum norm of the interaction Hamiltonian) in the energy eigenbasis. Tegmark further argues that this factorization corresponding to the energy eigenbasis is not the quasi-classical one despite maximum separability due to a crucial role played by the state.}.  In particular, we notice from Eqs. (\ref{Slin_oneterm}) and (\ref{Slin}) the presence of variance-like terms of the interaction Hamiltonian in the initial state. States that are more spread relative to terms in the interaction Hamiltonian (hence more variance) allow for more ways for the two subsystems to entangle and such features will play an important role in distinguishing the QC factorization. Interestingly, the self-Hamiltonian plays no role in entanglement production for initially unentangled states to $\Ottwo$. As we will see later, the self term is nevertheless important in determining the collimation of pointer observables under evolution, and will serve as an important feature of the QC factorization. 
 
 Not all unentangled states will allow for $\ddot{S}_{lin}(0) = 0$, even approximately, and only a special class of states for a given factorization will be robust to entanglement production. For an arbitrary factorization, there will not exist such entanglement-resilient states that do not get entangled (to $\Ottwo$) under evolution. When $\ddot{S}_{lin}(0) = 0$, for an arbitrary factorization where all $n_{int}$ terms are present in the interaction Hamiltonian without any constraints or relationship amongst different terms, each individual summand in Eq. (\ref{Slin}) will typically have to vanish separately, giving us $~ (n_{int})(n_{int} + 1)/2$ equations in the variables that make up the initial unentangled state $\ket{\psi(0)}_{A} \otimes \ket{\psi(0)}_{B}$. A generic unentangled state of this form has $(2d_{A} - 2)(2d_{B} -2) << (n_{int})(n_{int} + 1)/2$ real, free parameters (twice the dimension accounting for real coefficients; reduce two degrees of freedom, one due to normalization and one for the overall phase), hence forming an overdetermined set of equations. Only in very special cases, where quasi-classicality will be manifest will we see that many terms in $\intham$ will vanish having $\lambda_{\alpha} = 0$ or will conspire together to reduce/condense into familiar classical observables being monitored by other subsystems for there to exist robust, unentangled states that are resilient to entanglement production (and will serve as the pointer basis of the system). Such states will also be important for allowing decoherence to be an effective mechanism to suppress interference between superpositions of such pointer states.

\section{Predictability of Dynamics}
\label{sec:predictability}

\subsection{Pointer Observables and Predictable Diagonal-Sliding}
\label{sec:pointer}

The mere existence of a pointer observable consistently monitored by other subsystems is not enough for classical evolution of states starting with a peaked value of the observable. In addition to slow entanglement growth of initially unentangled pointer states, we must ensure that such states define a predictable variable that evolves classically. A possible measure for the predictability of an operator under evolution is the change in variance of the observable under an initial state with almost definitive value of the observable. Let us compute the time rate of change in variance of an observable $\Oa \in \mathcal{L}(\subA)$ under the evolution by $\ham$. {As one might expect, will see that the self-Hamiltonian $\ham_{A}$ becomes important in determining the how quickly the observable spreads. This feature will be tied with collimation properties of the $\ham_{A}$, which we discuss in detail in Section \ref{sec:collimation} in the context of conjugate variables.}

The variance of $\Oa$ as a function of time is defined as,
\begin{equation}
\label{Oa_var_1}
\Delta^{2}\Oa(t) = \trace\left( \oprho_{A}(t) \Oa^{2} \right) - \trace^{2}\left( \oprho_{A}(t) \Oa \right) \: .
\end{equation}
We will use the expression for $\oprho_{A}(t)$ to $\Ot$ from Eq. (\ref{rhoA_t_2}) since this is the lowest non-trivial order at which the effect of the Hamiltonian can be seen,
\begin{equation}
\oprho_{A}(t) = \opsig_{A}(t) - it \sum_{\alpha}\lambda_{\alpha} \lla \B_{\alpha} \rra_{0}  \lcb \Aa,\oprho_{A}(0) \rcb + \Ottwo \: ,
\end{equation}
which gives us,
\begin{equation}
\begin{split}
\label{var_int_1}
\trace\left( \oprho_{A}(t) \Oa^{2} \right) & = \lla \Oa^{2} \rra^{\mathrm{self}}_{t} - it \sum_{\alpha}\lambda_{\alpha} \lla \B_{\alpha} \rra_{0} \trace\left( \lcb \Aa,\oprho_{A}(0) \rcb \Oa^{2}\right) + \Ottwo \\
& = \lla \Oa^{2} \rra^{\mathrm{self}}_{t} - it \sum_{\alpha}\lambda_{\alpha} \lla \B_{\alpha} \rra_{0} \lla \lcb \Oa^{2},\Aa \rcb \rra_{0} + \Ottwo \: ,
\end{split}
\end{equation}
and similarly,
\begin{equation}
\label{var_int_2}
\trace\left( \oprho_{A}(t) \Oa \right) = \lla \Oa^{2} \rra^{\mathrm{self}}_{t} - it \sum_{\alpha}\lambda_{\alpha} \lla \B_{\alpha} \rra_{0} \lla \lcb \Oa,\Aa \rcb \rra_{0} + \Ottwo \: ,
\end{equation}
where the self-evolved variance $\left(\Delta^{2}\Oa\right)^{\mathrm{self}} $ is found similarly, depending on the self-Hamiltonian $\ham_{A}$,
\begin{equation}
\begin{split}
\label{var_self}
\left(\Delta^{2}\Oa\right)^{\mathrm{self}} (t) & = \trace\left( \opsig_{A}(t) \Oa^{2} \right) - \trace^{2}\left( \opsig_{A}(t) \Oa \right)  + \Ottwo  \\
& = \left(\Delta^{2}\Oa\right)_{0} - it  \biggl( \lla \lcb \Oa^{2},\ham_{A} \rcb \rra_{0} - 2 \lla \lcb \Oa,\ham_{A} \rcb \rra_{0} \lla\Oa \rra_{0}  \biggr) + \Ottwo \: .
\end{split}
\end{equation}
We can now put everything together to get the variance $\Delta^{2}\Oa(t)$ to $\Ot$,
\begin{equation}
\label{Oa_var_2}
\Delta^{2}\Oa(t) = \left(\Delta^{2}\Oa\right)^{\mathrm{self}} _{0} - it \sum_{\alpha}\lambda_{\alpha} \lla \B_{\alpha} \rra_{0} \biggl( \lla \lcb \Oa^{2},\Aa \rcb \rra_{0} - 2 \lla \lcb \Oa,\Aa \rcb \rra_{0} \lla\Oa \rra_{0}  \biggr) + \Ottwo \: .
\end{equation}

We can now obtain the leading order contribution to the time derivative of the variance that captures the contribution to various terms in the Hamiltonian,
\begin{equation}
\label{var_dot_1}
\begin{split}
\frac{d}{dt} \Delta^{2}\Oa(t) & = \biggl( \lla i \lcb \ham_{A},\Oa^{2} \rcb \rra_{0} - 2 \lla i \lcb \ham_{A}, \Oa \rcb \rra_{0} \lla\Oa \rra_{0}  \biggr) + \\
& \biggl( \lla i \lcb \sum_{\alpha}\lambda_{\alpha}\lla\B_{\alpha}\rra_{0}\Aa,\Oa^{2} \rcb \rra_{0} - 2 \lla i \lcb \sum_{\alpha}\lambda_{\alpha}\lla\B_{\alpha}\rra_{0}\Aa, \Oa \rcb \rra_{0} \lla\Oa \rra_{0}  \biggr)  + \Ot \: .
\end{split}
\end{equation}
The spreading of the variance depends on terms which resemble those in the Heisenberg equation of motion of the observable $\Oa$ (and its square) under evolution by both the self-Hamiltonian $\ham_{A}$ and relevant terms in $\intham$. 
 
Let us now analyze this variance change for the case where the interaction Hamiltonian $\intham$ in the chosen factorization admits a consistent pointer observable (in the QML) satisfying Eq. (\ref{Zurek_commutativity_QML}), in which case $\lcb \hat{f}(\Oa) , \Aa \rcb \approx 0 \: \: \forall  \: \: \alpha$ for any function $\hat{f}(\Oa)$ depending only on $\Oa$. For such a pointer observable, the time derivative of the variance from Eq. (\ref{var_dot_1}) simplifies and depends only on self-dynamics governed by $\ham_{A}$,
\begin{equation}
\label{var_dot_QML}
\frac{d}{dt} \Delta^{2}\Oa(t)  =  \lla i \lcb \ham_{A},\Oa^{2} \rcb \rra_{0} - 2 \lla i \lcb \ham_{A}, \Oa \rcb \rra_{0} \lla\Oa \rra_{0}   + \Ot \qquad \mathrm{for} \: \: \lcb \Oa , \Aa \rcb \approx 0 \: .
\end{equation}
For the pointer observable $\Oa$ to offer a predictable variable, it should obey $\frac{d}{dt} \Delta^{2}\Oa(t)  << 1$ for initial states that are peaked around some eigenvalue of $\Oa$. Having states as peaked superpositions of the pointer states instead of exact eigenstates fits in well with the idea of ``predictability sieve" \`a la Zurek \cite{Zurek:1994zq}: while the pointer observable is chosen using the compatibility criterion with $\intham$ as seen in Eq. (\ref{Zurek_commutativity_QML}), the most robust states (under the full Hamiltonian $\ham$) will have a small width instead of being exact eigenstates due to the effects of the (systematically smaller) self-Hamiltonian (in the QML). Such peaked states, for a predictable $\Oa$, will not spread much, and offer candidates for classical states that evolve primarily under the action of the self-Hamiltonian $\ham_A$. 

The reduced density matrix of $\subA$ in such a pointer basis will be mostly diagonal (due to decoherence, as discussed in Section \ref{sec:decoherence_feature}), and a peaked state of $\Oa$ will slide along the diagonal under self-dynamics \cite{tegmark2015consciousness}. This ``diagonal sliding" feature can also be seen from the expression for $\rhodot(t)$ from Eq. (\ref{rhoAdot_final}), where the diagonal entries of the decoherence term $\mathcal{D}(\oprho_{A}(t))$ of Eq. (\ref{decoherence_term}) in the pointer basis $\{\ket{s_{j}} \} $ vanish identically, and the diagonal entries in $\rhodot(t)$ in the pointer basis evolve as,
\begin{equation}
\label{rhodot_diagonal}
\left[ \frac{d}{dt } \oprho_{A}(t)\right]_{jj} = \biggl(-i \lcb \ham_{A}(t) , \oprho_{A}(t) \rcb_{jj}\biggr) + \Ottwo \: ,
\end{equation}
since even the interaction pieces from the effective self-Hamiltonian also vanish in the pointer basis (see Appendix \ref{app:non-generic} for details), $\braket{a_{j} | \lcb\Aa , \oprho_{A}(t)\rcb | a_{j}} \equiv \lcb\Aa , \oprho_{A}(t)\rcb_{jj} = 0$. Thus, these diagonal terms evolve under the action of the self-Hamiltonian and dictate the diagonal sliding of the density matrix in the pointer basis once it has decohered. 

Based on this motivation, we would like to capture predictability of the pointer observable to be used in conjunction with the entropy growth rate from Section \ref{subsec:min_entropy} to quantify the classicality of the system in question.
One can try $d \left(\Delta^{2} \Oa\right)/dt$ from Eq. (\ref{var_dot_QML}) as a measure of the predictability of the pointer observable, but such a quantity will not be a good homogeneous measure on the same footing as a dimensionless entropy like $S_{lin}$. This is because from the point of view of constructing an algorithm, we want to take into account both low entanglement growth and predictability of pointer observables to determine the QC factorization.

\subsection{Pointer Entropy}

To discuss an entropy measure that captures essentially the same physics as $d \left(\Delta^{2} \Oa\right)/dt$, we define a \emph{Pointer Entropy} as the second order $(q=2)$ Tsallis entropy of the probability distribution given by $\oprho_{A}(t)$ in the basis of an observable $\Oa$. 
\begin{equation}
\label{Spointer_define}
S_{pointer}(t) \: = \: 1 - \sum_{j = 1}^{d_{A}} p^{2}_{j}(t) \: ,
\end{equation}
where $p_{j}(t)$ is the probability distribution defined by,
\begin{equation}
\label{p_j_define}
p_{j}(t) = \Tr_{A} \biggl( \oprho_{A}(t) \ket{a_j}\bra{a_j} \biggr) \equiv  \Tr_{A} \biggl( \oprho_{A}(t) \hat{O}_{j} \biggr) = \braket{a_{j} | \oprho_{A}(t) | a_{j} }\: ,
\end{equation}
where $\{\ket{a_{j}}\}$ is the set of eigenstates of $\Oa$, and $\hat{O}_{j} \equiv \ket{a_{j}}\bra{a_j}$. Our goal is to be able to compare pointer entropy with linear entanglement entropy of Eq. (\ref{Slin}). As we saw in Section \ref{subsec:min_entropy}, a quantifier for entanglement robustness of unentangled states is $\ddot{S}_{lin}(0)$, and on similar lines we would like to use $\ddot{S}_{pointer}(0)$ as a quantifier for the rate of spread of the pointer observable. Following the discussion of Section \ref{sec:predictability} above, the reader should think of $\Oa$ as being the pointer observable of the system in the quasi-classical factorization, or else, if one is in a generic factorization, $\Oa$ should be thought of as the closest notion of a pointer observable (which we define precisely in Section \ref{sec:algorithm} below), which would best satisfy Zurek's commutativity criterion of Eq. (\ref{Zurek_commutativity_QML}) of being a pointer observable. Unlike arbitrary factorizations, for the case of the QC factorization, $\Oa$ would be the actual pointer observable and hence, would have predictable dynamics. 

$S_{pointer}$ is an entirely information-theoretic construction and is based on the probability distribution of $\oprho_{A}(t)$ in the basis of $\Oa$. It is insensitive to any ordering structure of eigenvalues and peaked states in this space. $S_{pointer}$ measures how far the spread of the probability distribution is from being completely certain, but does not capture its variance structure pertaining to a certain set of eigenvalues. Fortunately, for the class of states we are considering \textit{i.e.} those initial states which are peaked superpositions around some eigenvalue of $\Oa$ (and hence, are initially predictable states of $\Oa$), changes in $S_{pointer}$ correlate with a change in the variance $d \left(\Delta^{2} \Oa\right)/dt$ of the state itself.
\begin{figure}[h!]
\includegraphics[width= \textwidth]{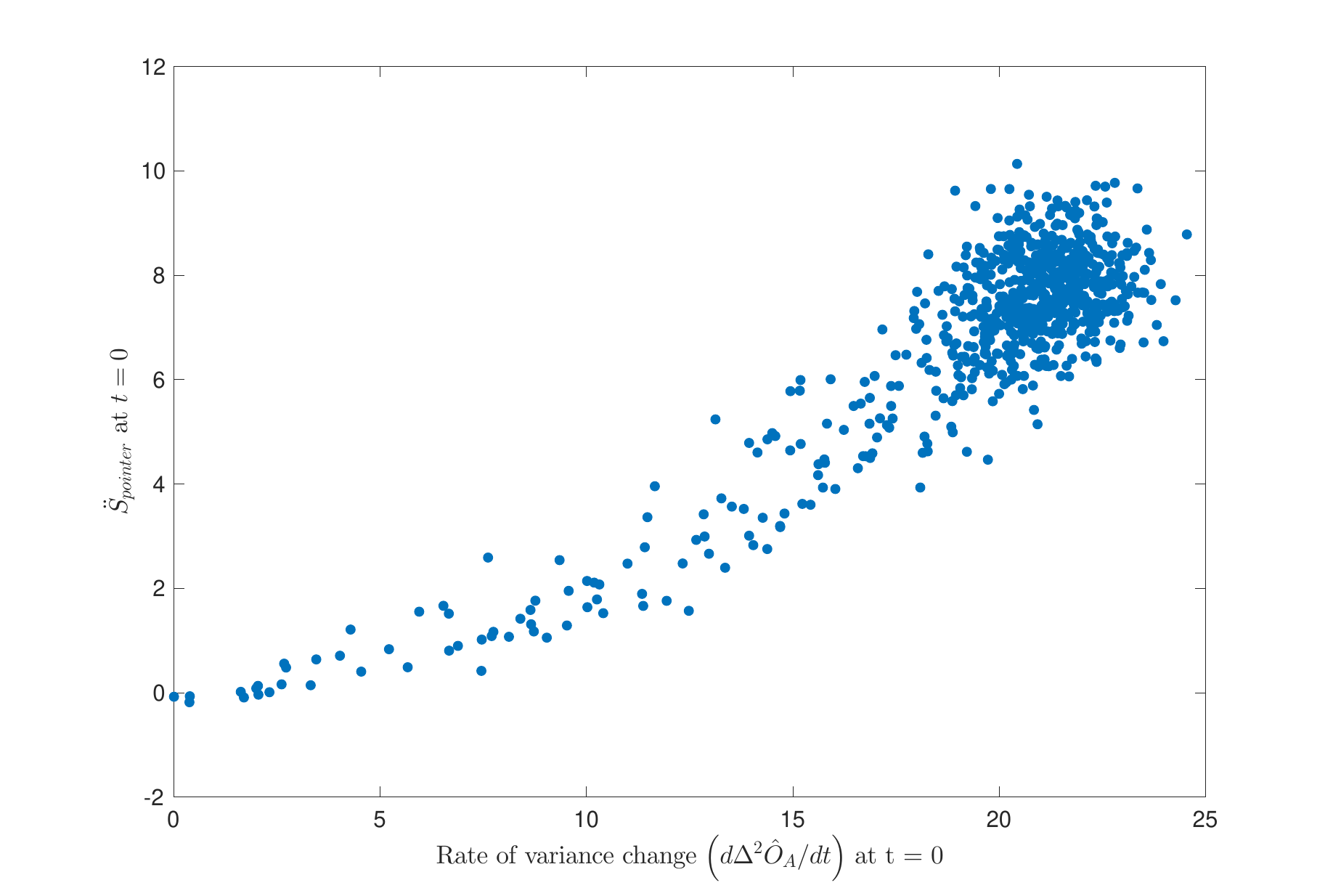}
\caption{Plot showing correlation between pointer entropy growth $\ddot{S}_{pointer}(0)$ and the rate of change of variance of the pointer observable for a peaked state in pointer eigenspace. Each point in the plot represents a different Hamiltonian. We kept $\Oa$ fixed (as the position operator, the finite-dimensional version of which is defined in Section \ref{sec:conjugate} below) and changed the self-Hamiltonian $\ham_{A}$ and computed the correlation for a peaked state in $\opphi_{A}$ eigenspace in a Hilbert space of $\Dim \subA = 27$.}
\label{schwinger_pointer}
\end{figure}

To further understand this point, let us focus on the QC factorization (while working in the QML) where $\Oa$ is the consistent pointer observable which satisfies Zurek's commutativity criterion of Eq. (\ref{Zurek_commutativity_QML}).  In Appendix \ref{app:pointer_ent_QML}, we compute $\ddot{S}_{pointer}(0)$ for this case when $\Oa$ is the pointer observable in the QC factorization, and we obtain,
\begin{equation}
\begin{split}
\ddot{S}_{pointer}(0) = 2\sum_{j=1}^{d_{A}} \lla \lcb \Oj,\ham_{A}\rcb \rra^{2}_{0} +
2\sum_{j=1}^{d_{A}}\left( p_{j}(0) \lla\Oj \ham^{2}_{A} + \ham^{2}_{A}\Oj - 2\ham_{A}\Oj\ham_{A}\rra_{0} \right) + \\
2\sum_{j=1}^{d_{A}}\left( p_{j}(0) \sum_{\alpha}\lambda_{\alpha} \lla \B_{\alpha} \rra_{0} \lla\lcb\Oj,\lcb\ham_{A},\Aa\rcb\rcb \rra_{0}  \right)\:  .
\end{split}
\end{equation}
Based on the properties of the self-Hamiltonian, the rate of change of variance $d \left(\Delta^{2} \Oa\right)/dt$ strongly correlates with the rate of growth of pointer entropy $\ddot{S}_{pointer}(0)$ for peaked initial states, and we plot this result in Figure \ref{schwinger_pointer} (details in caption). We will make further contact with features of the self-Hamiltonian that lead to small changes in $S_{pointer}$ (as in the QC factorization) in Section \ref{sec:collimation} when we discuss the role of emergent classical conjugate variables.

\section{The Quantum Mereology Algorithm}
\label{sec:algorithm}

Given a Hilbert space and a Hamiltonian, how does one sift through Hilbert space factorizations and pick out the one corresponding to the QC decomposition? This section aims to use the features described in Section \ref{sec:qc_factorization} as pointers to outline an algorithm that quantifies the quasi-classicality of each factorization and uses this to pick out the one in which the QC features are most manifestly seen. We will do this for the bipartite case we have focused on in this paper. {We will be focusing on the case where the dimension of each subsystem is fixed, \textit{i.e.} $d_A$ and $d_B$ are specified a priori (not necessarily prime). The goal is to find the factorization which exhibits manifestly classical dynamics corresponding to the given, fixed sizes of the two subsystems. The question of iterating over possible dimension split $D = D_{A}d_{B}$ is left for future investigation.} As we have seen, features like existence of low entropy states and robustness against entanglement production, non-generic decoherence and predictability of pointer observables are highly special and particular to the QC factorization and will not be seen in other, arbitrary factorizations. Hence, for an algorithm that sifts through Hilbert space factorizations, we need to identify a homogeneously defined quantity for each factorization that would be extremized for the QC factorization. We will use the $S_{lin}$ computation from Section \ref{subsec:min_entropy} and predictability of the pointer observable from Section \ref{sec:pointer} to identify such a quantity. 

\subsection{Candidate Pointer Observables}

As we have seen, low entropy states obtained as eigenstates of a consistent pointer observable that are resilient to entanglement production ($\ddot{S}_{lin} = 0$ to $\Ottwo $ in our calculation) are highly non-generic; they are a special feature of the QC factorization. To belabor this point a little more, consistent pointer observables of the form $\Oa \otimes \hat{O}_{B}$ with $\lcb \intham, \Oa \otimes \hat{O}_{B} \rcb \approx 0$ do not exist generically exist. This motivates us to define a \emph{Candidate Pointer Observable} (CPO) for an arbitrary factorization $\tdec$ that can serve as a proxy for the pointer observable by being the closest observable consistently monitored by the environment. Of course, for the QC decomposition $\tdec_{QC}$, the CPO coincides with the pointer observable, and for other factorizations away from the QC, the CPO will introduce a ``penalty" term in our measure of predictability and robustness of classical states in the factorization.

{The CPO $\hat{O}_{CPO}$ is the non-trivial observable (\textit{i.e.} not the null operator, or a multiple of the identity operator) most compatible with $\intham$ that has a factorizable form akin to a typical pointer observable,
 \begin{equation}
 \hat{O}_{CPO} \equiv  \hat{\tilde{O}}_{A} \otimes \hat{\tilde{O}}_{B},
 \end{equation} 
 for {some} operators $\hat{\tilde{O}}_{A} \in \subA$ and $ \hat{\tilde{O}}_{B} \in \subB$. Thus, $\hat{O}_{CPO}$ serves as the best (with regards to a norm measure, which we choose to be the Frobenius norm denoted by a subscript ''$F$'') product operator that is compatible with $\intham$, and can therefore serve as a proxy/best possible notion of a consistent pointer observable,
\begin{equation}
\label{CPO}
\hat{O}_{CPO} = \hat{\tilde{O}}_{A} \otimes \hat{\tilde{O}}_{B} \: \: \mathrm{such \: that \:}  \left|\left| \lcb \intham, \hat{O}_{CPO} \rcb  \right|\right|_{F} \:  \: \mathrm{is \: minimized \:}.
\end{equation}
In the QC factorization, $\hat{O}_{CPO}$ will coincide with an actual pointer observable which is consistently monitored by the environment. To obtain the CPO, and to ensure it is non-trivial, we can parametrize $\hat{\tilde{O}}_{A}$ and $\hat{\tilde{O}}_{B}$ using the hermitian, traceless Generalized Gell-Mann matrix basis (details in Appendix \ref{GGMM}),
\begin{equation}
\label{eq:OA_OB_GGM}
\hat{\tilde{O}}_{A} = \sum_{j = 1}^{d^{2}_{A} - 1} a_{j} \hat{\Lambda}^{(A)}_{j} \: , \: \:\:\:\:\:\:\:\: \:\:\:\:\:\:\:
\hat{\tilde{O}}_{b} = \sum_{k = 1}^{d^{2}_{B} - 1} b_{k} \hat{\Lambda}^{(B)}_{k} \: ,
\end{equation}
for $a_{j}, b_{k} \in \mathbb{R}$ (to ensure hermiticity of the CPO), with at least one element of the set $\{a_{j} \}$, and one element of the set $b_{k}$ being non-zero. To ensure uniformity in searching for the CPO, we also demand that $\left|\left|\hat{\tilde{O}}_{A}   \right|\right|_{F} \: = \: \left|\left|\hat{\tilde{O}}_{B}   \right|\right|_{F} \: = \: 1$ which constrains the values the sets $\{a_{j} \}$ and $\{b_{k} \}$ can take. Since there is no support of the identity element in the matrix expansion of Eq. (\ref{eq:OA_OB_GGM}), and the Generalized Gell-Mann matrices are hermitian and traceless, we ensure that the obtained CPO is non-trivial and therefore can neither be the null operator nor any multiple of the identity. The minimization of Eq. {\ref{CPO}} can then be carried out, albeit numerically if necessary.}
\\
\\

The next thing to focus on is the kind of state we will be using to quantify the quasi-classicality of a given factorization $\tdec$. As we have seen, peaked states of a consistent pointer observable can serve as good candidates for studying predictability, and in the correct limit be identified as classically predictable states.
Following this motivation, we can construct states that represent peaked states of the CPO $\hat{\tilde{O}}_{A}$ on $\subA$, 
\begin{equation}
\ket{\psi_{j}(0)}_{CPO} = \ket{\tilde{\psi} (0)}_{A} \otimes \ket{\tilde{\psi} (0)}_{B}. 
\end{equation}
They represents an initially predictable state for our subsystem $\subA$ having a definite value of the candidate pointer observable. One possible prescription for the state $\{\ket{\tilde{\psi}_{j} (0)}_{A}\}$ is to construct a peaked state around an eigenstate of $\hat{\tilde{O}}_{A}$, and take the state on $\subB$ to be a uniform superposition of all eigenstates of $\hat{\tilde{O}}_{B}$ to represent a ready state for the candidate environment $\subB$. 

{To measure the robustness to entanglement following the discussion in Section \ref{sec:robustness}, one can now compute $\ddot{S}_{lin}$ for the state $\ket{\psi(0)}_{CPO}$,
\begin{equation}
 \ddot{S}_{lin}(0)  =  \bigg(\frac{d^{2}}{dt^{2}}S_{lin}(\oprho_{A}(t)) \bigg)\bigg|_{t=0} \: ,
\end{equation}
(by using Eq. \ref{Slin} for the explicit expression) which will serve as a measure of the entanglement resilience of low entropy states in the decomposition $\tdec$.} For the particular case of the QC factorization $\tdec_{QC}$, we will find $\ddot{S}_{lin}$ for $\ket{\psi(0)}_{CPO}$ to  {vanish} (or even approximately so) since the state will correspond to one constructed out of a consistent pointer observable that is robust to entanglement production under evolution. For other factorizations $\ddot{S}_{lin} \neq 0$ will serve as a penalty quantifier, with higher the value of $\ddot{S}_{lin}$, the more non classical the factorization. 

{The other side of the story comes from predictability of dynamics as discussed in Section \ref{sec:predictability}. Since we established the connection between pointer entropy and rate of variance change established for the case of a consistent pointer observable in the QC factorization, we can broaden now our computation of $\ddot{S}_{pointer}(0)$ to a more general situation that will be useful in quantifying a predictability measure for the Candidate Pointer Observable (CPO).
Using the general expressions for $\oprho_{A}(t)$ and $d\oprho_{A}/dt$ from Eqs. (\ref{oprhoA_Ot}) and (\ref{rhoAdot_final}), we find,
\begin{equation}
\label{Spointerdoubledot_general}
\ddot{S}_{pointer}(0) = -2\sum_{j=1}^{d_{A}}\left(\dot{p}^{2}_{j}(0) + p_{j}(0) \ddot{p}_{j}(0)\right) \: .
\end{equation}
We refrain from writing the full general expression here to avoid unnecessary clutter, but the important thing to remember is that for an arbitrary factorization, $\ddot{S}_{pointer}(0)$ for a peaked initial state for the Candidate Pointer Observable will serve as a quantifier for predictability of the candidate. As one goes closer to the QC factorization $\tdec_{\theta}$, the CPO matches with a true pointer observable, and thus becomes highly predictable. In other factorizations, the value of $\ddot{S}_{pointer}(0)$ will typically be higher as a penalty for the factorization not admitting a good pointer observable.
}

\subsection{The Algorithm}
\label{subsec:algo}

We can now summarize the Quantum Mereology Algorithm, which sifts through various bipartite factorizations of Hilbert space and searches for the QC factorization. The algorithm will extremize an entropic quantity built from a combination of $S_{lin}$ and $S_{pointer}$ to pick out the QC factorization which shows both features of robustness and predictability as outlined in Section \ref{sec:qc_factorization}. 

For an arbitrary decomposition $\tdec$, {of a Hilbert space with fixed dimensions $d_A$ and $d_B$ of the bipartite subsystems,}
\begin{enumerate}
\item{Find the Candidate Pointer Observable $\hat{O}_{CPO} = \hat{\tilde{O}}_{A} \otimes \hat{\tilde{O}}_{B} $ from Eq. (\ref{CPO}), which is the best tensor product observable compatible with the interaction Hamiltonian. }
\item{Construct a set of states that represent peaked states of the CPO $\hat{\tilde{O}}_{A}$ on $\subA$, $\ket{\psi_{j}(0)}_{CPO} = \ket{\tilde{\psi}_{j} (0)}_{A} \otimes \ket{\tilde{\psi} (0)}_{B}$. They represent an initially predictable state for our subsystem under consideration $\subA$ having a definite value of the candidate pointer observable. To ensure quasi-classical conditions hold for all pointer states in the QC factorization, construct $d_A$ number of such states, labeled by $j = 1,2,\ldots,d_{A}$. One possible prescription for these $d_A$ states $\{\ket{\tilde{\psi}_{j} (0)}_{A}\}$ is to construct peaked states around each eigenstate of $\hat{\tilde{O}}_{A}$, and take the state on $\subB$ to be a uniform superposition of all eigenstates of $\hat{\tilde{O}}_{B}$ in each case to represent a ready state for the candidate environment $\subB$. }
\item{For each of these states, $\ket{\psi(0)}_{CPO}$, compute $\ddot{S}_{lin}(0)$ and $\ddot{S}_{pointer}(0)$ from Eqs. (\ref{Slin}) and (\ref{Spointerdoubledot_general}), respectively. These are measures of the quasi-classicality of the factorization. Lower $\ddot{S}_{lin}(0)$ indicates a factorization whose pointer states (from the CPO) are robust to entanglement production, and lower $\ddot{S}_{pointer}(0)$ indicates a factorization which preserves predictability of classical states under evolution.}
\item{Define \emph{Schwinger Entropy} (here, its second derivative) as follows,
\begin{equation}
\label{S_schwinger}
\ddot{S}_{Schwinger} = \mathrm{max} \biggl(\ddot{S}_{lin}(0), \ddot{S}_{pointer}(0) \biggr) \: .
\end{equation}
Average over the $d_A$ states from the eigenstates CPO to obtain the value of $\ddot{S}_{Schwinger}$ for the given factorization. We choose to label this quantity as Schwinger entropy to serve as a reminder that we are  using Schwinger's unitary basis (from the GPOs) to define our construction in a finite-dimensional context.}
\item{Find the factorization that minimizes $\ddot{S}_{Schwinger}$. This will be the quasi-classical factorization.}
\end{enumerate}
Using this procedure, one can sift through different factorizations of Hilbert space, and pick out the one that shows features of robustness and predictability of pointer states as the quasi-classical factorization. In the next section, we will make contact with finite-dimensional conjugate variables, and understand features of the Hamiltonian which allow us to interpret the conjugates as classical positions and momenta.

\section{Emergence of Classical Conjugate Variables} 
\label{sec:conjugate}

{Before turning to an explicit example of the algorithm in action, we would like to examine the way in which the classical limit arises according to our procedure.
At an abstract level, a classical system is described by points in a symplectic manifold that evolve according to a vector field derived from a Hamiltonian function.
But in practice, real-world systems have additional features such as locality, which work to distinguish position variables from momentum variables.
Here we argue that this feature can be understood in terms of how factorizations of Hilbert space are chosen to recover classical behavior.
In particular we introduce the idea of the collimation of an operator, which helps characterize the self-Hamiltonian of subsystems that exhibit classical behavior.}

\subsection{Finite-Dimensional Conjugate Variables}

Classical mechanics is formulated in phase space, with conjugate position and momentum variables.
For quantum mechanics in infinite-dimensional Hilbert spaces, we can define corresponding quantum operators, subject to the Heisenberg canonical commutation relations (CCR). 
Since we are explicitly focusing on finite-dimensional Hilbert spaces, we will use generalized Pauli operators (which find their algebraic roots in the generalized Clifford algebra) to provide us with finite-dimensional conjugate variables that obey the CCR in the infinite-dimensional limit. 
We will then use these to define what we call the ``collimation'' of an operator, an important notion that characterizes how the action of an operator on a state induces a spread in eigenspace.

We explain the basics of generalized Pauli operators (GPOs) in Appendix~\ref{app:gca}.
The essential point is that we can construct Hermitian conjugate operators $\hat{q}$ and $\hat{p}$ that match onto position- and momentum-like operators in the infinite-dimensional limit.
To do this we introduce two unitary operators $\A$ and $\B$ that will generate the GPO algebra.
On a Hilbert space of dimension $d < \infty$, they obey the Weyl braiding relation,
\begin{equation}
\label{weylbraid}
\hat{A}\hat{B} = \omega^{-1}\hat{B}\hat{A} \: ,
\end{equation}
where $\omega = \exp\left(2 \pi i /d\right)$ is the $d$-th primitive root of unity, and are sometimes referred to as ``Clock'' and ``Shift'' operators in the literature.
Then the conjugate variables are defined via
\begin{equation}
\label{phipidef}
\A \equiv \exp{(-i \alpha \oppi)} \: , \: \: \: \: \B = \exp{(i \beta \opphi)} \: ,
\end{equation}
where $\alpha$ and $\beta$ are non-zero real parameters that set the scale of the eigenspectrum of the operators $\opphi$ and $\oppi$ with a cyclic structure. For concreteness, we take the dimension to be an odd integer, $d = 2l + 1$ for some $l \in \mathbb{Z}^{+}$.

The set of $N^2$ linearly independent unitary matrices $\{B^{b}A^{a} | b,a = -l , (-l+1),\cdots,0,\cdots,(l-1),l \}$, which includes the identity for $a = b = 0$, form a unitary basis for  $\mathcal{L}(\hs)$. Schwinger \cite{Schwinger570} studied the role of such unitary basis, hence this operator basis is often called {Schwinger's unitary basis}. Any operator $\hat{M}  \in \mathcal{L}(\hs)$ can be expanded in this basis,
\begin{equation}
\label{Mexpansion}
\hat{M} = \sum_{b,a = -l}^{l} m_{ba} \B^b \A^a \: .
\end{equation}
Since from the structure of the GPO algebra we have $\Tr  \left[ \left( \B^{b'} \A^{a'} \right)^{\dag}  \left( \B^{b} \A^{a}\right) \right]= d \: \delta_{b,b'} \delta_{a,a'}$, we can invert Eq.~(\ref{Mexpansion}) to get the coefficients $m_{ba}$ as,
\begin{equation}
\label{mba_coeff}
m_{ba} = \frac{1}{d} \Tr \left[ \A^{-a} \B^{-b} \hat{M} \right] \: .
\end{equation} 
The GPO generator $\A$ corresponds to a unit shift in the eigenstates of $\opphi$, and $\B$ generates unit shifts in the eigenstates of $\oppi$; hence, a basis element  $B^{b}A^{a}$ generates $a$ units of shift in eigenstates of $\opphi$ and $b$ units in eigenstates of $\oppi$, respectively (up to overall phase factors). 

For an operator $\M$ that is Hermitian $\M^{\dag} = \M$, we get a constraint on the expansion coefficients, $\omega^{-ba} m^{*}_{-b,-a} = m_{ba}$, which implies $|m_{ba}| = |m_{-b,-a}|$ since $\omega = \exp{\left( 2 \pi i/(2l+1) \right)}$ is a primitive root of unity. 
The coefficients $m_{ba}$ are a set of basis-independent numbers that quantify the spread induced by the operator $\hat{M}$ along each of the conjugate variables $\opphi$ and $\oppi$. To be precise, $|m_{b,a}|$ represents the amplitude of $b$ shifts along $\oppi$ for an eigenstate of $\oppi$ and $a$ shifts along $\opphi$ for an eigenstate of $\opphi$ . The indices of $m_{ba}$ run from $-l, \cdots, 0, \cdots, l$ along both conjugate variables and thus, characterize shifts in both increasing $(a$ or  $b > 0)$ and decreasing $(a$ or  $b < 0)$ eigenvalues on the cyclic lattice. The action of $\hat{M}$ on a state depends on details of the state, and in general will lead to a superposition in the eigenstates of the chosen conjugate variable as our basis states, but the set of numbers $m_{ba}$ quantify the spread along conjugate directions by the operator $\hat{M}$ independent of the choice of state. The coefficient $m_{00}$ accompanies the identity $\eye$, and hence corresponds to no shift in either of the conjugate variables. 

From $m_{ba}$, which encodes amplitudes of shifts in both $\opphi$ and $\oppi$ eigenstates, we would like to extract profiles which illustrate the spreading features of $\M$ in each conjugate variable {separately}. Since the coefficients $m_{ba}$ depend on details of $\M$, in particular its norm, we define normalized amplitudes $\tilde{m}_{ba}$ for these shifts,
\begin{equation}
\label{mba_norm}
\tilde{m}_{ba} = \frac{m_{ba}}{\sum_{b',a' = -l}^{l} |m_{b'a'}|} \: .
\end{equation}
Then we define the {$\opphi$-shift profile} of $\M$ by marginalizing over all possible shifts in $\oppi$,
\begin{equation}
\label{ushiftham}
m^{(\phi)}_{a} \: = \: \sum_{b = -l}^{l} |\tilde{m}_{ba}| \: = \: \frac{\sum_{b = -l}^{l} |m_{ba}|}{\sum_{b',a' = -l}^{l} |m_{b'a'}|}  \: ,
\end{equation}
which is a set of $(2l+1)$ positive numbers, normalized under $\sum_{a=-l}^{l} m^{(\phi)}_{a} = 1$, characterizing the relative importance of $\M$ spreading the $\opphi$ variable by $a$ units, $a = -l,\cdots,0,\cdots,l$. Thus, $\M$ acting on an eigenstate of  $\opphi$, say $\ket{\phi = j}$, will in general, result in a superposition over the support of the basis of the $\opphi$ eigenstates $\{ \ket{\phi = j + a \: (\mathrm{mod} \: l)} \}  \: \: \forall \: a$, such that the relative importance (absolute value of the coefficients in the superposition) of each such term is upper bounded by $m^{(\phi)}_{a}$. 
 
Let us now quantify this spread by defining the collimation for each conjugate variable. Consider the {$\phi$-shift profile} first. Operators with a large $m^{(\phi)}_a$ for small $|a|$ will have small spread in the $\opphi$-direction, while those with larger $m^{(\phi)}_{a}$ for larger $|a|$ can be thought of connecting states further out on the lattice for each eigenstate. Following this motivation, we define the  $\phi$-collimation $C_{\phi}$ of the operator $\M$ as,
\begin{equation}
\label{Su}
C_{\phi}(\M) = \sum_{a =-l}^{l} m^{(\phi)}_{a} \exp{\left(- \frac{|a|}{2l + 1} \right)} \: .
\end{equation}
The exponential function suppresses the contribution of large shifts in our definition of collimation. There is some freedom in our choice of the decay function in our definition of collimation, and using an exponential function as in Eq. (\ref{Su}) is one such choice. Thus, an operator with a larger $C_{\phi}$ is highly collimated in the $\opphi$-direction and does not spread out eigenstates with support on a large number of basis states on the lattice. 

On similar lines, one can define the {$\pi$-shift profile} for $\M$ as,
\begin{equation}
\label{vshiftham}
m^{(\pi)}_{b} \: = \: \sum_{a = -l}^{l} |\tilde{m}_{ba}| \: = \: \frac{\sum_{a = -l}^{l} |m_{ba}|}{\sum_{b'a' = -l}^{l} |m_{b'a'}|}  \: ,
\end{equation}
and a corresponding {$\pi$-collimation} $C_{\pi}$ with a similar interpretation as the $\opphi$-case,
\begin{equation}
\label{Sv}
C_{\pi}(\M) = \sum_{b =-l}^{l} m^{(\pi)}_{b} \exp{\left(- \frac{|b|}{2l + 1} \right)} \: .
\end{equation}
Operators such as $\M \equiv \M(\oppi)$ that depend on only one of the conjugate variables will only induce spread in the $\opphi$ direction, since they have $m_{b,a} = m_{0,a} \delta_{b,0}$, hence they possess maximum $\pi$-collimation, $C_{\pi}(\M) = 1$, as they do not spread eigenstates of $\oppi$ at all. 

 \begin{figure}[t]
\includegraphics[width=\textwidth]{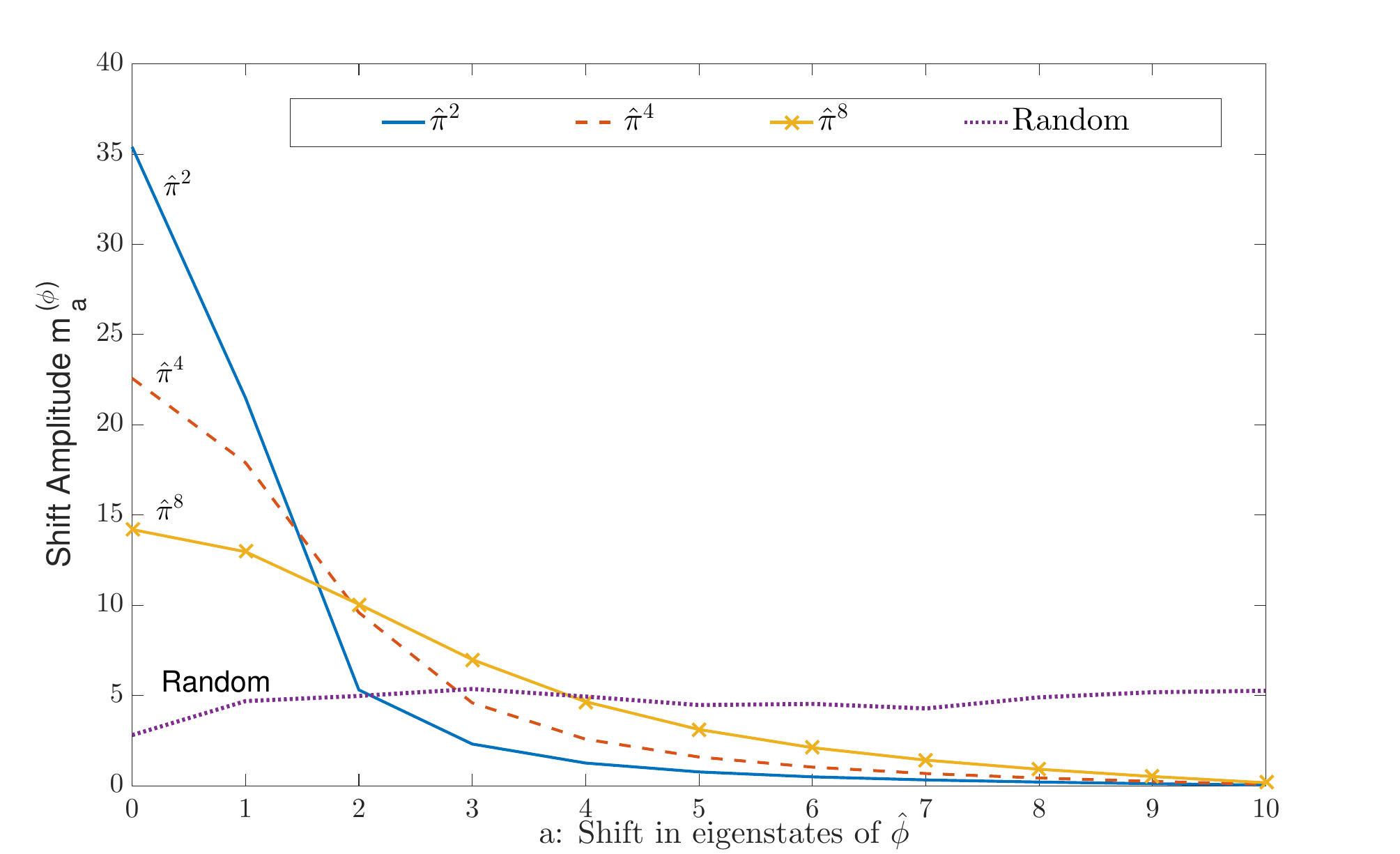}
\caption{Plot showing $\opphi$-shift profiles of various powers of $\oppi$. The quadratic operator $\oppi^{2}$ has the most collimated profile, implying that this operator does the least to spread the state in the conjugate direction. Also plotted is the profile for a random Hermitian operator, for which the spread is approximately uniform.}
\label{schwinger_profiles}
\end{figure}

While the maximum value of $C_{\pi}(\M(\oppi))$ can be at most unity, one can easily see that the Hermitian operator,
\begin{equation}
\M(\oppi) = \frac{A + A^{\dag}}{2} = \frac{\exp{\left(- i \alpha \oppi\right)} + \exp{\left(i \alpha \oppi\right)}}{2} = \cos\left( \alpha \oppi \right) = \eye - \frac{\alpha^{2} \oppi^{2}}{2} + \frac{\alpha^{4} \oppi^{4}}{4} - \cdots \: ,
\end{equation}
has the least non-zero spread along the $\opphi$ direction: it connects only $\pm 1$ shifts along eigenstates of $\opphi$ and hence has highest (non-unity) $\phi$-collimation $C_{\phi}(\M)$. 
Thus, one can expect operators which are quadratic in conjugate variables are highly collimated. This will connect to the fact that real-world Hamiltonians include terms that are quadratic in the momentum variables (but typically not higher powers) and will help explain the emergence of classicality: it is Hamiltonians of that form that have high position collimation, and therefore induce minimal spread in the position variable. 
\begin{figure}[h]
\includegraphics[width=\textwidth]{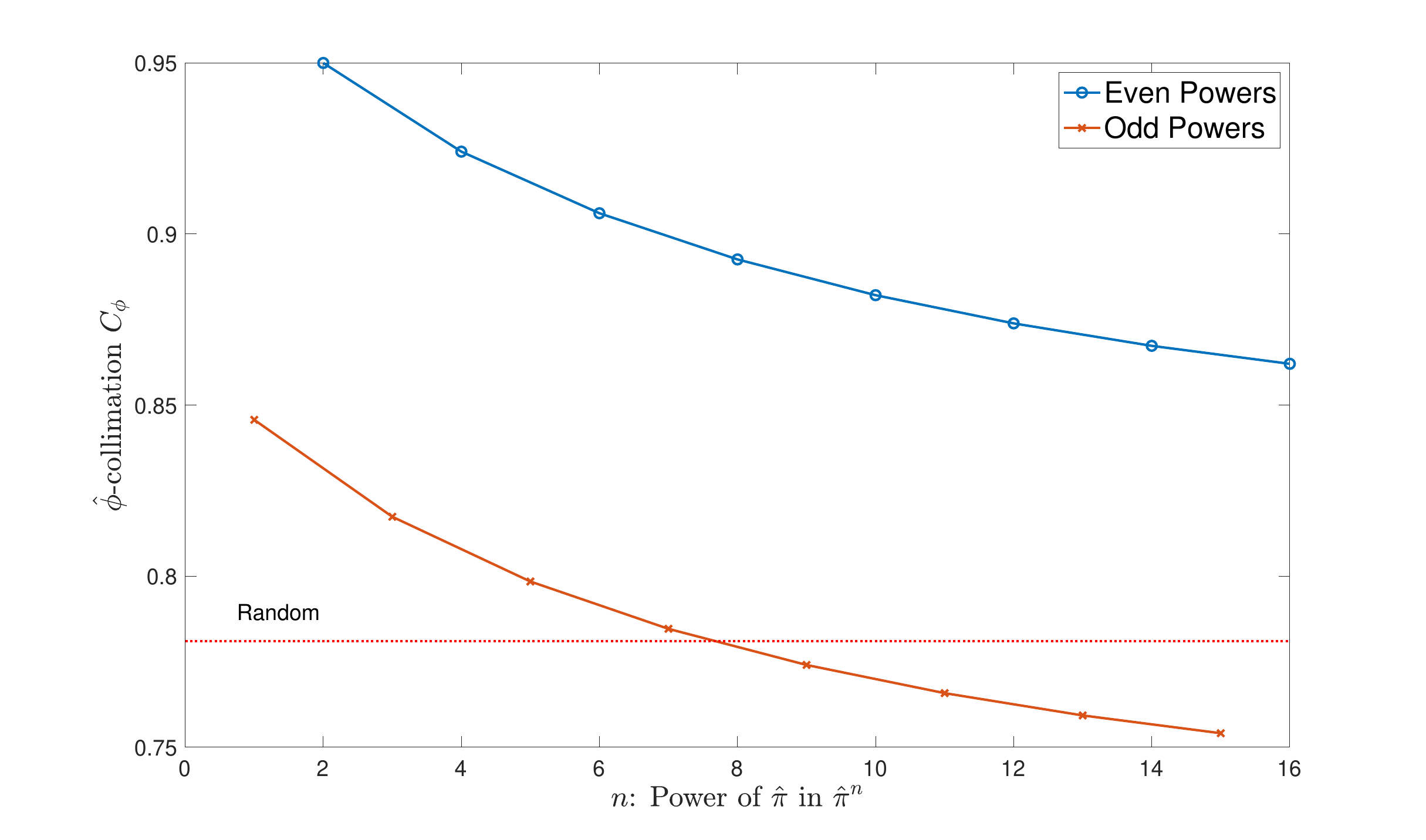}
\caption{$\phi$-collimation of various powers of $\oppi$. Even powers are seen to have systematically larger values of $\phi$-collimation. Also plotted for comparison is a line marking the $\phi$-collimation of a random Hermitian operator.}
\label{schwinger_powers}
\end{figure}

The quadratic operator $\oppi^{2}$ has higher $\phi$-collimation than any other integer\footnote{There is a difference between odd and even powers of $\oppi$, with even powers systematically having larger collimations than the odd powers. This is because odd powers of $\oppi$ no have support of the identity $\eye$ term in the Schwinger unitary basis expansion (and hence have $m_{00} = 0$), and having an identity contribution boosts collimation since it contributes to the highest weight in $C_{\phi}$ by virtue of causing no shifts.} power $\oppi^{n}, n \geq 1 \: , \: n \neq 2$. In Figure (\ref{schwinger_profiles}), we plot the $\phi$-shift profiles for a few powers of $\oppi$ and it is explicitly seen that quadratic $\oppi^{2}$ has the least spreading and hence is most $\opphi$-collimated, values for which are plotted in Figure (\ref{schwinger_powers}). Note that due to the symmetry $|m_{b,a}| = |m_{-b,-a}|$, we only needed to plot the positive half for $a > 0$, which captures all the information about the spread. Also, for comparison, we also plot the $\phi$-spread and the $\opphi$-collimation of a {random} Hermitian operator (with random matrix elements in the $\opphi$ basis); such operators spread states almost evenly and thus have low values of collimation. 

\subsection{Operator Collimation, Locality, and the Self-Hamiltonian}
 \label{sec:collimation}
 
Typically, one begins with a notion of classical subsystems, then defines the Hamiltonian for these systems based on classical energy functions, and proceeds to quantize. In non-relativistic quantum mechanics, the self terms usually go as $~ \hat{p}^2/2m + \hat{V}(\hat{q})$ for canonically conjugate operators $\hat{p}$ and $\hat{q}$. Interaction terms usually depend on one of the conjugate variables, usually the position of each subsystem.  

For each subsystem one can associate a set of finite-dimensional conjugate operators from the Generalized Pauli Operators.
For our bipartite split $\hs = \subA \otimes \subB$, we have conjugate operators $\{\opphi_{A},\oppi_{A}\} \in \mathcal{L}(\subA)$ and $\{\opphi_{B},\oppi_{B} \} \in \mathcal{L}(\subB)$. For arbitrary factorizations, these GPO-based conjugate variables will not correspond to physical position and momentum variables;  only in a quasi-classical decomposition would the identification $\oppi \equiv \hat{p}$ and $\opphi \equiv \hat{q}$ be appropriate. 

The conjugate variables can be used to define the Schwinger Unitary Basis \cite{Schwinger570}, and hence we can write self terms in the Hamiltonian $\ham$ from (\ref{H_bipartite}) in terms of these conjugates,
\begin{equation}
\label{ops_self_conjugate}
\ham_{A} \equiv \ham_{A}\left( \oppi_{A},\opphi_{A} \right) \: \: \: \: \ham_{B} \equiv \ham_{B}\left( \oppi_{B},\opphi_{B} \right) \: ,
\end{equation}
and the interaction term can be written as,
\begin{equation}
\label{ops_int_conjugate}
\intham \equiv \intham\left(\oppi_{A},\opphi_{A},\oppi_{B},\opphi_{B} \right) \: = \: \sum_{\alpha}  \lambda_{\alpha} \biggl( \Aa\left(\oppi_{A},\opphi_{A}\right) \otimes \B_{\alpha}\left(\oppi_{B},\opphi_{B}\right) \biggr) \: .
\end{equation}

Before we explicitly discuss the idea of collimation and the role it plays in emergent quasi-classicality, let us comment on the functional form of $\intham$ for there to exist a robust pointer observable as described in the previous Section \ref{sec:qc_factorization}. Since $\opphi_{A}$ and $\oppi_{A}$ do not commute, for there to exist a  compatible pointer observable monitored consistently by other subsystems, we would demand that interaction terms $\Aa$ depend only on one such conjugate variable, say $\opphi_{A}$. In many physical cases, the interaction term is the position of the subsystem under consideration, as that is the quantity that is monitored by the environment, since interactions are local in space. Under these conditions, the pointer observable can be identified as $\Oa \equiv \Oa(\opphi_{A})$, depending only on one conjugate variable.

\begin{figure}[h]
\includegraphics[width= \textwidth]{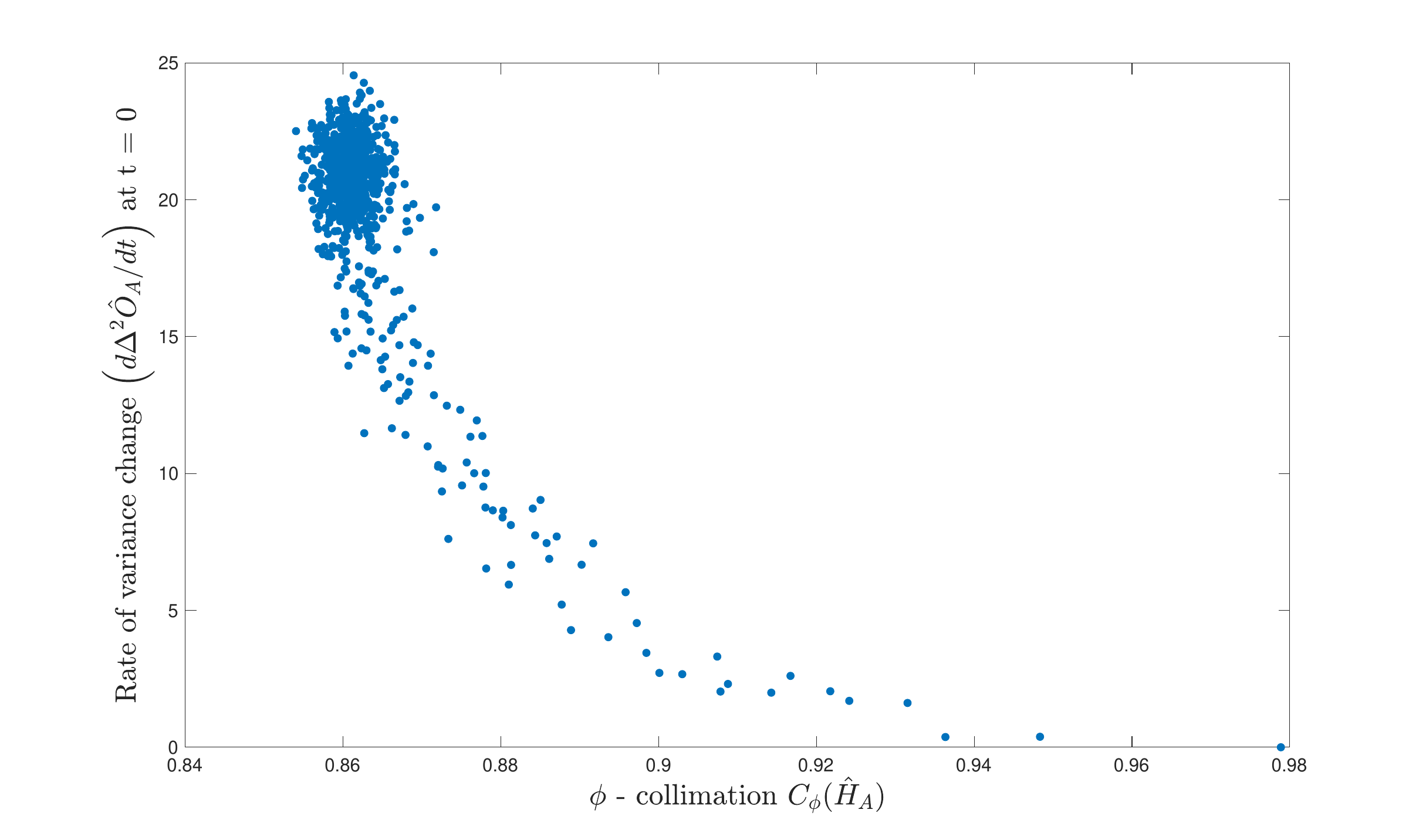}
\caption{Plot showing correlation between rate of change of variance of the pointer observable $d \left(\Delta^{2} \Oa\right)/dt$ and collimation $C_{\phi}(\ham_{A})$ of the self-Hamiltonian $\ham_{A}$. More collimated self terms do not spread states much in the conjugate directions and correspondingly induce a small change in the variance of the consistent pointer observable that depends on one of these conjugate variables. In this example, we kept $\Oa$ fixed at $\Oa \equiv \opphi_{A}$ and changed the self-Hamiltonian $\ham_{A}$ and computed the correlation for a peaked state in $\opphi_{A}$ eigenspace in a Hilbert space of $\Dim \subA = 27$.}
\label{schwinger_vardot}
\end{figure}

 Let us see how the idea of predictability connects with features of the self-Hamiltonian. From Eq. (\ref{var_dot_QML}), we see that the rate of change of variance of the pointer observable depends, in addition to the state at $t = 0$, on the Heisenberg equation of motion for $\Oa(\opphi_{A})$ under the self-Hamiltonian $\ham_{A}$. Thus, it can be expected that self terms $\ham_{A}$ that are collimated in the $\opphi_{A}$ variable will spread states less rapidly under time evolution and keep the change of variance of $\Oa$ small. They therefore offer a predictable interpretation to $\Oa$. This can be seen in the following example. We keep fixed the pointer observable $\Oa \equiv \opphi_{A}$ and vary the self-Hamiltonian, and for each choice of the self-Hamiltonian we compute the time derivative of $\Delta^2 \Oa$ from Eq. (\ref{var_dot_QML}) for an initial state that is a peaked Gaussian profile in $\opphi_{A}$ states, representing a peaked wavepacket.  In Figure (\ref{schwinger_vardot}), we plot these results and see that high $\phi$-collimation $C_{\phi}(\ham_{A})$ inversely correlates with the variance change of the pointer observable. Therefore, evolving under a highly $\phi$-collimated self-Hamiltonian, peaked states in pointer space have a smaller rate of change of variance of the pointer observable $\Oa$.
 
One can then interpret the results of Figs. \ref{schwinger_pointer} and \ref{schwinger_vardot} together to correlate the pointer entropy growth with the collimation of the self-Hamiltonian, which will play a crucial role in determining how fast the pointer entropy spreads out, thus quantifying the predictability of the pointer observable. Self-Hamiltonians with a higher collimation will induce smaller spread and hence a slower growth in pointer entropy (and rate of change of variance) for peaked pointer states.

Note the different roles played by collimation and locality. 
In quantum field theories or lattice theories, we can factor Hilbert space into sets of degrees of freedom located in small regions of space.
Spatial locality then implies that the interaction Hamiltonian takes a $k$-local form, where each factor interacts directly with only its neighboring factors; cf. Eq.~(\ref{M_int}).
For our purposes we can turn this around, looking for factorizations in which interactions are $k$-local, which is a necessary requirement for the emergence of spatial locality \cite{cotler2019locality}.
Collimation, by contrast, is an important feature of the self-Hamiltonian.
In order to recover familiar classical behavior, we require that pointer observables evolve in relatively predictable ways, rather than being instantly spread out over a wide range of values.

\subsection{Classical Dynamics}
\label{sec:classical_dynamics}

Besides the existence of predictable pointer observables, the other feature we require for quasi-classical behavior is that conjugate ``position'' and ``momentum'' operators, or some generalization thereof, approximately obey the corresponding classical Heisenberg equations of motion.

As was shown in Ref \cite{Singh:2018qzk}, and this argument can be easily extended for multi-partite systems, the equations of motion for the conjugate operators $\opphi_{A}$ and $\oppi_{A}$ for some subsystem $\subA$ can be found to be,
\begin{equation}
\label{Veom_finite}
\frac{d}{dt} \oppi = i \left[ \ham, \oppi_{A} \right]  = -  \widehat{\left(\frac{\partial H}{\partial \phi_{A}}\right)}  + \sum_{n = 3}^{\rm{odd}} \frac{i^n}{n!} \alpha^{n-1} \left[ \underline{\oppi_{A}} , \ham \right]_{n} \: ,
\end{equation}
where we have defined $\left[ \underline{\oppi_{A}} , \ham \right]_{n}$ as the $n$-point nested commutator in $\oppi_{A}$,
\begin{equation}
\label{npointcommutator}
\left[ \underline{\oppi_{A}} , \ham \right]_{n} = \left[\oppi_{A},\left[\oppi_{A},\left[\oppi_{A} \cdots \: (n\mathrm{\ times)}, \ham \right]\cdots\right]\right] \: .
\end{equation}
The corresponding equation for $\opphi_{A}$ can be found on similar lines,
\begin{equation}
\label{Ueom_finite}
\frac{d}{dt} \opphi_{A} = i \left[ \ham, \opphi_{A} \right]  =  \widehat{\left(\frac{\partial H}{\partial \pi_{A}}\right)} + \sum_{n = 3}^{\rm{odd}} \frac{i^n}{n!} \beta^{n-1} \left[ \underline{\opphi_{A}} , \ham \right]_{n} \: .
\end{equation}

In the infinite-dimensional limit, we take $l \gg 1$, and $\alpha$ and $\beta$ are taken to be infinitesimal but obeying $\alpha \beta d = 2\pi$ to recover the Heisenberg CCR in an appropriately understood sense \cite{doi:10.1080/09500349014551931}.\footnote{A finite-dimensional Hilbert space, however large, cannot be isomorphic to an infinite-dimensional one. Due to this, the $d\gg1$ limit of the finite-dimensional commutation $\lcb \opphi, \oppi \rcb$ will not formally give the Heisenberg CCR. This is due to the fact that unitary translation operators are cyclically closed in the finite-dimensional construction, because of which their hermitian generators are trace-class, unlike their infinite-dimensional counterparts on $\mathbb{L}^{2}(\mathbb{R})$. One manifestation of this is that the trace of the finite-dimensional commutator will identically vanish, whereas it is ``infinite'' in the infinite-dimensional case. Many of the involved sums become highly oscillatory, and it is important to deal with them appropriately. However, as was shown in Ref. \cite{FLORATOS199735}, a large number of eigenvalues of the finite-dimensional commutator $\lcb \opphi, \oppi \rcb$ approach $i$ in the $d\gg1$ limit, and in this sense, if one further restricts attention to states which do not have substantially large support (on the eigenbasis of either $\opphi$ or $\oppi$) for the cyclic closure effects to matter, then the large-dimension limit can be safely taken. In our analysis, we are predominantly focusing on peaked states of a pointer observable for which these effects would not matter, and hence we are able to interpret our results in the large-dimension limit.}  
For our analysis, we take $\alpha = \beta = \sqrt{{2\pi}/{d}}$ since there is no distinguishing $\opphi$ and $\oppi$ at the level of the algebra. Their symmetry is broken in their physical interpretation as position and momentum, respectively due the different roles they play in the Hamiltonian.
In this limit the equations of motion simplify to resemble Hamilton's equations,
\begin{equation}
\label{Veom_infinite}
\frac{d}{dt} \oppi_{A} = i \left[ \ham, \oppi_{A} \right]  = - \widehat{\left(\frac{\partial H}{\partial \phi_{A}}\right)} \: ,
\end{equation}
and,
\begin{equation}
\label{Ueom_infinite}
\frac{d}{dt} \opphi_{A} = i \left[ \ham, \opphi_{A} \right]  =  \widehat{\left(\frac{\partial H}{\partial \pi_{A}}\right)} \: ,
\end{equation}
where $\ham$ is the Hamiltonian for the entire Hilbert space. Even though in the large-dimension limit these resemble classical equations of motion, they are inherently quantum mechanical equations for operators in $\mathcal{L}(\hs)$. Additional features have to be imposed for the conjugate variables $\opphi_{A}$ and $\oppi_{A}$ to serve as classical conjugate variables. 

These equations serve as classical evolution equations when we consider peaked states of the pointer observable $\Oa$ that would depend on only one of the conjugate variables, say $\Oa \equiv \Oa(\opphi_{A})$. Peaked states in $\Oa$ eigenspace can be candidates for classical evolution since they can obey the Ehrenfest theorem when one takes expectation value of Eqs. (\ref{Veom_infinite}) and (\ref{Ueom_infinite}) by pulling in the expectation into the Hamiltonian, for example, 
\begin{equation}
\label{ehrenfest1}
\lla \widehat{\left(\frac{\partial H}{\partial \pi_{A}}\right)} \rra \to \left(\frac{\partial \lla \ham \rra}{\partial \pi_{A}}\right) \: \mathrm{for \:  peaked \:  states \: in \: pointer \: observable \: space.} \: 
\end{equation}
The condition for persistence of such classical states obeying classical equation of motion will be to have low spreading of the variance of such a peaked state, which as we saw corresponds to a highly collimated self-Hamiltonian. Thus, under the criterion of there existing a predictable and consistent pointer observable (from $\intham$ in the Quantum Measurement Limit) that depends on one of the conjugate variables and a collimated self-Hamiltonian, we would be able to identify the conjugate variables $\opphi_{A}$ and $\oppi_{A}$ (from the GPO algebra) with {classical conjugate variables}. While one can always define conjugates, the existence of classical ones corresponding to our familiar notion of position and momenta are highly non-generic and connect to predictability features in the Hamiltonian.

\section{Example}
\label{subsec:algo_action}

We now demonstrate the algorithm with a simple example where we recover the quasi-classical factorization by sifting through different factorizations of Hilbert space and selecting the one which minimizes Schwinger entropy for candidate classical states.
Let us take our complete quantum system to be described by two harmonic oscillators, coupled together (interacting) by their positions in the quasi-classical factorization. We take both these oscillators to have the same mass $m$ and same frequency $\omega$, and thus having their respective self-Hamiltonians,
\begin{equation}
\ham_{A} = \frac{\oppi^{2}_A}{2m} + \frac{1}{2}m\omega^{2}\opphi^{2}_{A} \: ,
\end{equation}
\begin{equation}
\ham_{B} = \frac{\oppi^{2}_B}{2m} + \frac{1}{2}m\omega^{2}\opphi^{2}_{B} \: .
\end{equation}
The interaction term is modeled as oscillator $A$'s position $\opphi_{A}$ coupled to the position $\opphi_B$ of oscillator $B$ with an interaction strength $\lambda$,
\begin{equation}
\ham_{int}  = \lambda \left(\opphi_{A} \otimes \opphi_{B}\right) \: .
\end{equation}

This conventional way of writing the model makes physical sense to us, and implies a corresponding factorization of Hilbert space.
As we now show, this choice matches our above criteria for a quasi-classical factorization as elaborated in Sections \ref{sec:robustness} and \ref{sec:predictability}. 
The interaction Hamiltonian in the QC factorization takes the simple form $\ham_{int} = \lambda(\A \otimes \B )$ that is compatible with having low entropy pointer states robust to entanglement under evolution. The pointer observable of subsystem $\subA$ under consideration is the position $\opphi_A$ of that oscillator. The self-Hamiltonian is highly collimated with respect to $\opphi_A$, as can be seen by the quadratic power of $\oppi_A$ in $\ham_{A}$. We choose values of the parameters $m,\omega,\lambda$ such that we are in the quantum measurement limit (QML) where the interaction term dominates.

 \begin{figure}
\includegraphics[width=\textwidth]{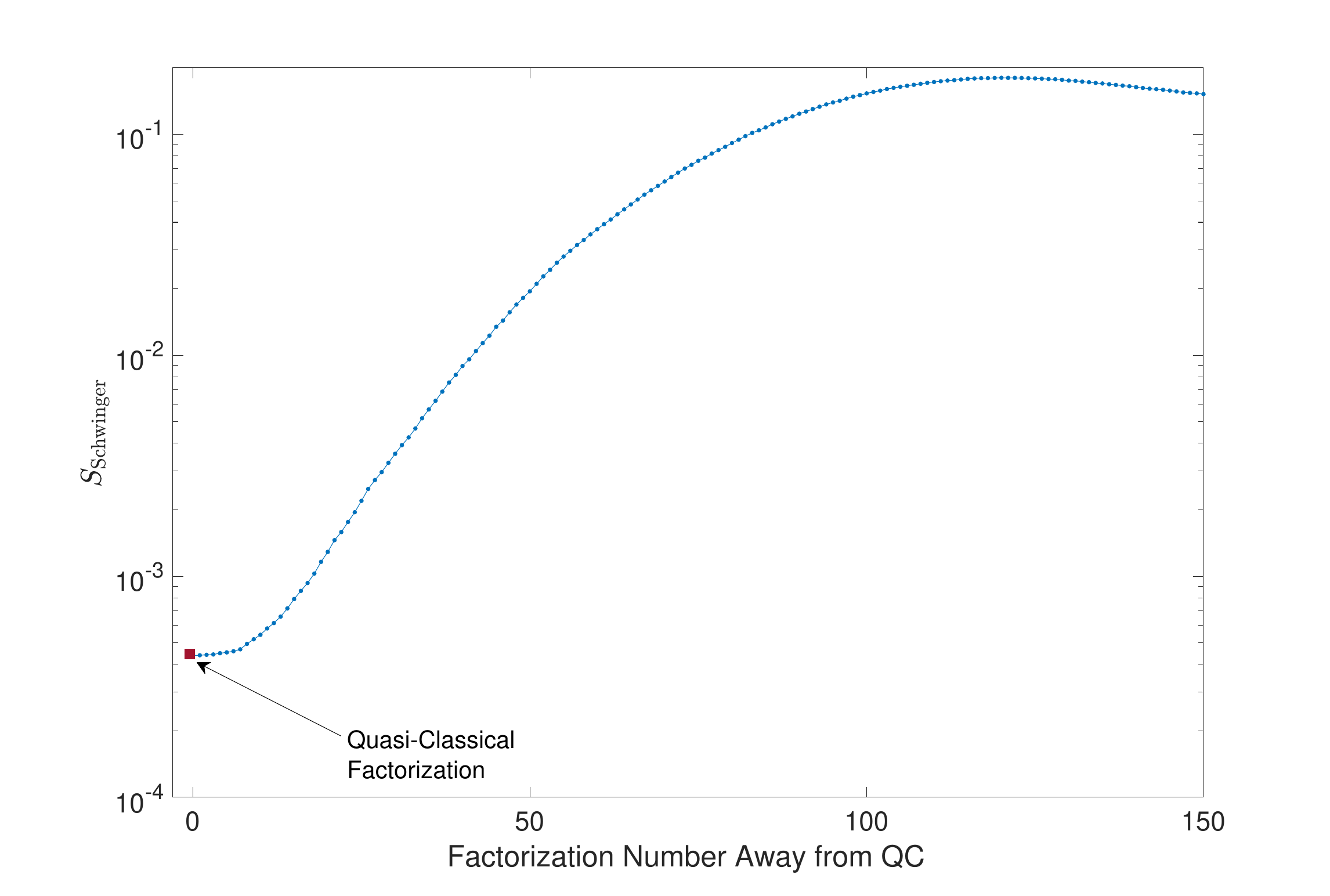}
\caption{Quantum Mereology Algorithm: sifting through different factorizations of Hilbert space to recover the QC factorization by minimization of the Schwinger entropy. In the QC factorization, the quantum system is described by two harmonic oscillators coupled by their positions. The quasi-classical factorization (the first factorization we begin with) is marked by a red square. Details in the text.}
\label{fig:mereo_1}
\end{figure}

We now demonstrate the Quantum Mereology algorithm by ``forgetting" that we start in the QC factorization, and try to recover it by sifting through factorizations and select the QC one by minimization of the Schwinger entropy. We change factorizations by introducing incremental, random perturbations away from the identity operator (by making the parameters $\{\theta\}$ non-zero in Eq. \ref{U_expansion}) to construct the global unitary transformation $\tilde{U}(\theta)$. 
Since we are focusing on the quantum measurement limit, we make sure perturbations do not get large enough so as to break down the assumption of applicability of the QML regime (for example, a factorization change to make the two oscillators completely decoupled would no longer be in the QML, and hence we do not focus on such factorizations in this paper). For each factorization, while the total Hamiltonian is left invariant, the form of the self and interaction terms are altered. We run the Quantum Mereology Algorithm as outlined in Section \ref{subsec:algo} with choosing eigenstates of the CPO $\hat{\tilde{O}}_{A}$ as our peaked initial, low entropy states (one could construct peaked superpositions too which does not alter the results).

 \begin{figure}
\includegraphics[width=\textwidth]{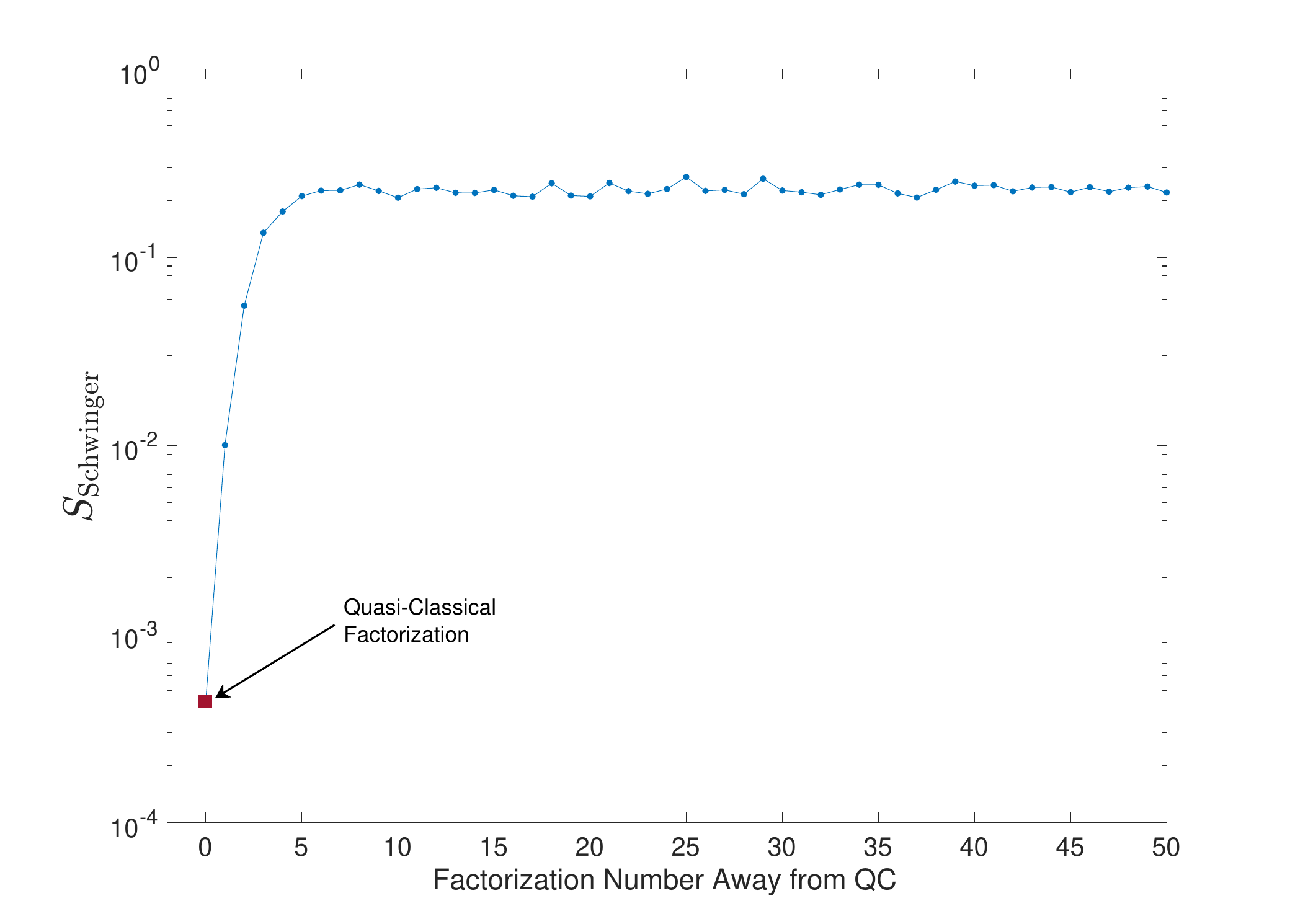}
\caption{Quantum Mereology Algorithm: a run similar to figure \ref{fig:mereo_1} sifting through different factorizations of Hilbert space but this time with larger, successive perturbations away from the identity operator to generate the unitary transformation $\Lambda$. In the QC factorization, which minimizes the Schwinger entropy, the quantum system is described by two harmonic oscillators coupled by their positions. Factorizations away from QC quickly saturate to generic, large values of Schwinger entropy. The quasi-classical factorization (the first factorization we begin with) is marked by a red square. Details in the text.}
\label{fig:mereo_2}
\end{figure}

 In figure \ref{fig:mereo_1}, we plot the Schwinger entropy for many factorizations the algorithm sifts through, beginning with the QC factorization and then scrambling away. Since we are focusing on small times, we evolve quantum states to a characteristic time of $t_{0} = 1/||\hat{H}||_{2}$ and use the Schwinger entropy $S_{\mathrm{Schwinger}}$ at $t = t_{0}$ as the representative measure of classicality, instead of explicitly computing the second derivative at $t=0$, $\ddot{S}_{\mathrm{Schwinger}}(t=0)$. This is done purely for convenience, and does not affect the interpretation of picking out the QC factorization since for small times, both the linear entanglement entropy and pointer entropy grow as $t^2$. It is seen that Schwinger entropy is minimized for the QC factorization, which exhibits features of both robustness and predictability. In figure \ref{fig:mereo_2}, we plot the results of a similar run but this time with larger, successive perturbations away from the QC factorization (while still being in the quantum measurement limit). While in figure \ref{fig:mereo_1}, we see a more gradual deviation from classicality, in \ref{fig:mereo_2}, there is more rapid growth and saturation of the Schwinger entropy to larger values which are characteristic of generic, non-classical factorizations.

\section{Discussion}

In this paper we have developed a set of criteria, and an associated algorithm, for starting with an initially featureless Hilbert space $\hs$ and Hamiltonian $\ham$, and factorizing $\hs$ into a system and environment, optimizing the extent to which the system exhibits quasi-classical behavior. 
The basic criteria we introduced were that system pointer states remain relatively robust against increasing entanglement with the environment, and that pointer observables evolve in relatively predictable ways.
Both notions were quantified in terms of entropy: the linear entropy for entanglement robustness, and pointer entropy for predictability.
Useful factorizations are those that minimize the growth of both of these entropies, which we suggested combining into a single ``Schwinger entropy."

This work suggests a number of open questions and directions for future investigation.
Let us briefly note some of them:
\begin{itemize}
\item{While promising in principle, it is unclear how feasible our algorithm is in practice. Given nothing but the spectrum of a finite-dimensional Hamiltonian, it could take a long time to sift through the space of factorizations to find the one that minimizes the Schwinger entropy growth rate. It would be interesting to look for more computational feasible algorithms, even if only approximate. Schwinger entropy, as defined in Eq. (\ref{S_schwinger}), is a suggestive quantity which attempts to capture both of the classical features (of robustness and predictability) we want. Further work could also focus on refining the definition and purview of such a quantity.}
\item{We focused on how to factorize Hilbert space into system and environment, but ultimately we would want to continue to factorize the system into appropriate subsystems. We believe that the same basic strategy should apply, though locality and other considerations may come into play. It would also be interesting to look at an algorithm which sifts over different dimensions of the underlying Hilbert space sub-factors to study the effect of varying dimension associated with classical subsystems.}
\item{We looked exclusively at the Quantum Measurement Limit, in which the system is continuously monitored by the environment. The other extreme case of the Quantum Decoherence Limit is when the self-Hamiltonian dominates, and the pointer states are energy eigenstates of the self-Hamiltonian. We feel that the same essential concepts should apply, but it would be interesting to look at this case more explicitly.}
\item{The stability of the quasi-classical factorization is another interesting question to study. Do quasi-classical features stay preserved under infinitesimal perturbations of the factorizations or is quasi-classicality finely tuned? We expect classicality to be a robust feature enhanced by the existence of multiple subsystems each redundantly recording information about the others. This ties back into the idea of Quantum Darwinism \cite{riedel2012} and it would be interesting to investigate this question further. }
\end{itemize}

As we mentioned at the start, in typical laboratory situations the choice of how to factorize Hilbert space is fairly evident, and the question of mereology doesn't arise.
But as we consider more abstract theories, including those of emergent spacetime in quantum gravity \cite{Cao:2016mst, Cao:2017hrv}, our laboratory intuition may no longer be relevant, and an algorithm of the sort presented here can become important.
The separation into system and environment that we considered here may be related to how states are redundantly specified in a quantum error-correcting code \cite{Harlow:2016vwg,Cao:2017hrv}.
It is certainly a central concern of the program of reconstructing the quasi-classical world from the spectrum of the Hamiltonian \cite{Carroll:2018rhc,cotler2019locality}.
Regardless, it is important to understand in principle why we impose the structures on Hilbert space that we do.

\section*{Acknowledgments}
We would like to thank Anthony Bartolotta, Ning Bao, ChunJun (Charles) Cao, Aidan Chatwin-Davies, Jason Pollack and Jess Riedel for helpful discussions during the course of this project. We are also thankful to two anonymous referees for their comments to help us improve the manuscript. This research is funded in part by the Walter Burke Institute for Theoretical Physics at Caltech, by the Foundational Questions Institute, and by the U.S. Department of Energy, Office of Science, Office of High Energy Physics, under Award Number DE-SC0011632.

\appendix

\section{Bases and Factorizations}
\label{app:factor}

To factorize a Hilbert space of finite dimension $\Dim \hs = D < \infty$ is to express it as a tensor product of $N$ smaller factors,
\begin{equation}
\label{TF_1}
\hs \simeq \bigotimes_{\mu}^{N} \hs_{\mu} \: .
\end{equation}
The factors $\hs_{\mu}$ have dimensions $d_{\mu}$.
These need not be equal for all $\mu$, but their product must give the overall dimension, $\prod_{\mu}^{N} d_{\mu} = D$.

The most straightforward way to specify a factorization is in terms of a tensor-product basis that is adapted to it. For convenience we take all of our bases to be orthonormal.
In each factor $\hs_{\mu}$ we fix a basis,
\begin{equation}
\hs_{\mu} \simeq \mathrm{span}\{\ket{e^{(\mu)}_{i}}\} \: , \: \: \: i = 1,2,\ldots, d_{\mu} \: .
\end{equation}
We can then define basis vectors for $\hs$ as a whole by taking the tensor product of individual basis elements,
\begin{equation}
\label{TPB_elements}
\hs \simeq \bigotimes_{\mu = 1}^{N}\mathrm{span}\{\ket{e^{(\mu)}_{i}}\} \: .
\end{equation}
Of course such bases are highly non-unique; unitary transformations within each separate factor will leave the associated factorization itself unchanged.

In practice, one can construct different factorizations of $\hs$ by starting with some {reference} factorization and associated tensor-product basis, then performing a unitary transformation that mixes factors.
To implement the change in decomposition, we pick a special unitary matrix $\tilde{U} \in SU(D) \backslash \left( \bigotimes_{\mu=1}^{N} U(d_{\mu}) \right)$ that is characterized by $(D^2 - 1)$ real parameters $\{ \theta_{a} \: | \: a = 1,2,\cdots, (D^2 - 1) \} $ and has $D^2 - 1$ traceless, Hermitian generators $\{ \Lambda_{a}\ : | \: a = 1,2,\cdots, (D^2 - 1) \} $, which can be identified with the Generalized Gell-Mann matrices (GGMM). These GGMMs come in three groups: symmetric, anti-symmetric and diagonal matrices. In the notation where $E^{jk}$ is the $D \times D$ matrix with all zeros, except a $1$ in the $(j,k)$ location, the GGMMs have the following form, each identified with one of the $\Lambda_{a}$,
\begin{subequations}
\label{GGMM}
\begin{align}
\Lambda_{\rm{sym}}^{jk} = E^{kj} + E^{jk} \: \: ; \: \: 1 \leq j < k \leq D, \\    
\Lambda_{\rm{antisym}}^{jk} = -i\left( E^{jk} - E^{kj} \right)\: \: ; \: \: 1 \leq j < k \leq D, \\    
\Lambda_{\rm{diag}}^{l} = \sqrt{\frac{2}{l(l+1)}} \left(-l \: E^{l+1,l+1} + \sum_{j = 1}^{l}E^{jj} \right) \: \: ; \: \: 1 \leq l \leq D - 1 \: .
\end{align}
\end{subequations}
We work with special unitary instead of unitary since the global $U(1)$ phase is irrelevant to the physics of factorization changes and now we can express the factorization change unitary $\tilde{U} (\theta)$ as,
\begin{equation}
\label{U_expansion}
\tilde{U}(\theta) = \exp{\left( \sum_{a = 1}^{D^2 - 1} \theta_{a} \Lambda_{a} \right)} \: ,
\end{equation}
and factorization changes can be implemented on the reference decomposition. 

In light of this parametrization, let us label decompositions by the set of parameters $\{\theta \}$, which are used to implement the factorization change relative to the reference decomposition $\{0\}$. This notation will help us succinctly show dependence of various quantities on the factorization of Hilbert space. Product states in the old tensor-product basis (such as basis states in this factorization) will now be entangled in the new global basis identified with a new tensor product structure. Generally, operators that are local in their action to a certain sub-factor in a given decomposition such as $\hat{O}_{\nu} \equiv \eye \otimes \eye \otimes \cdots \otimes \hat{o}_{\nu} \otimes \cdots  \otimes \eye$, which act non-trivially only on the $\nu$-th subsystem, will generically act on more than one sub-factor in a different factorization. Locality of operators is a highly factorization-dependent statement and it has been shown \cite{cotler2019locality} that most tensor factorizations of Hilbert space for a given Hamiltonian do not look local and the existence of dual local descriptions is rare and almost unique.

In a given decomposition, any operator $\hat{M} \in \mathcal{L}(\hs)$, the space of linear operators on $\hs$ can then be naturally decomposed as,
\begin{equation}
\label{M_self_int}
\hat{M} =  \left(\frac{m_{0}}{D}\right) \eye \:  + \:  \sum_{\mu = 1}^{N} \hat{M}^{\mathrm{self}}_{\mu}  \: + \: \hat{M}_{\mathrm{int}} \: ,
\end{equation}
where $m_{0} = \Tr(\hat{M})$ is the trace of $\hat{M}$, the operator $\hat{M}^{\mathrm{self}}_{\mu}$ is the (traceless) term that acts locally only on the $\hs_{\mu}$ sub-factor and an interaction term, also traceless, connecting different sub-factors $\hat{M}_{\mathrm{int}}$. The interaction term can be decomposed further as a sum of $n$-point interactions,
\begin{equation}
\label{M_int}
\hat{M}_{\mathrm{int}} = \sum_{n = 2}^{N} \left( \sum_{\mu_{1} > \mu_{2} > \cdots > \mu_{n}} \hat{M}_{\mathrm{int}}(\mu_{1},\mu_{2},\cdots,\mu_{n}) \right) \: ,
\end{equation}
where, $\hat{M}_{\mathrm{int}}(\mu_{1},\mu_{2},\cdots,\mu_{n})$ is a term connecting sub-factors labeled by $\mu_{1}, \mu_{2},\cdots,\mu_{n}$. Any traceless, local term $\hat{M}_{\mu}$ that acts on a single sub-factor $\hs_{\mu}$ can be expanded out in the basis of Generalize Gell-Mann operators $\hat{\Lambda}^{\mu}_{a}$ with $a = 1,2,\cdots,(d^{2}_{\mu} - 1)$, which are $(d^{2}_{\mu} - 1)$ traceless, Hermitian generators of the $SU(d_{\mu})$,
\begin{equation}
\label{M_mu_expansion}
\hat{M}_{\mu} \equiv \eye \otimes \eye \otimes \cdots \otimes \hat{M}_{\mu} \otimes \cdots \otimes \eye = \eye \otimes \eye \otimes \cdots \otimes \sum_{a = 1}^{d^{2}_{\mu} - 1} m_{a} \hat{\Lambda}_{a}  \otimes \cdots \otimes \eye \: .
\end{equation}
In general, any operator $\hat{M}$ can also be decomposed in the canonical operator basis formed from the defining tensor-product basis,
\begin{equation}
\label{M_TPB_expansion}
\hat{M} = \sum_{i,j = 1}^{D} m_{ij} \ket{e_{i}}_{TPB} \bra{e_{j}} \: .
\end{equation}
Such expansions do not necessarily show the locality and interaction terms explicitly, but in the preferred, semi-classical decomposition, one would be able to arrange them in the form of familiar semi-classical terms in which features like robustness, quasi-separability and decoherence will be manifest.

\section{Evolution of the Linear Entropy}
\label{sec:low_entropy}
 
In this section we calculate the evolution of the linear entropy $S_{lin}$ to $\Ottwo$, leading to Eq. (\ref{Slin}).
Using the Zassenhaus expansion, which is a corollary of the Baker-Campbell-Haursdorff (BCH) lemma, one can separate the sum in the time evolution exponential $\hat{U}(t)$ as,
\begin{equation}
\Uop(t) = \exp{\left( - i (\selfham + \intham) t \right)} \: ,
\end{equation}
\begin{equation}
\Uop(t) = \exp{\left(- i \intham t \right)} \exp{\left(- i \selfham t \right)} \exp{\left(- \frac{(-it)^2}{2}  \left[ \intham,\selfham \right] \right)} \exp{\left( \mathcal{O}(t^3) \right)} \: .
\label{zassenhaus1}
\end{equation}
We can move the $\exp{\left(- \frac{(-it)^2}{2}  \left[ \intham,\selfham \right] \right)}$ past the $\exp{\left(- i \selfham t \right)}$ term to the left since the commutator we pick up is $\mathcal{O}(t^3)$ as can be explicitly checked by use of the Zassenhaus expansion again to get,
\begin{equation}
\label{zassenhaus2}
\Uop(t)  = \exp{\left(- i \intham t \right)}  \exp{\left(- \frac{(-it)^2}{2}  \left[ \intham,\selfham \right] \right)}  \exp{\left(- i \selfham t \right)}\exp{\left( \mathcal{O}(t^3) \right)} \: .
\end{equation}
Further one can see that the first two pieces involving $\intham$ and $\left[ \intham,\selfham \right]$ in the above equation Eq. (\ref{zassenhaus2}) can be combined into a sum of a single exponential since the non-commuting pieces will be $\mathcal{O}(t^3)$, and this gives us a succint expression for $\Uop(t)$ to $\mathcal{O}(t^2)$,
\begin{equation}
\label{Utime2}
U(t) = \exp{\left( - i \E(t) t \right)} \: \exp{\left( - i \selfham t \right)} + \mathcal{O}(t^{3})  \: ,
\end{equation}
where,
\begin{equation}
\label{Eop}
\E(t) \equiv \intham + \frac{i t}{2} \left[ \intham,\selfham \right] \: .
\end{equation}

Taking $\Uop(t)$ from Eq. (\ref{Utime2}), the time evolved state can be written as $\oprho(t) = \Uop(t) \oprho(0) \Uop^{\dag}(t)$ to $\mathcal{O}(t^2)$. Let us define {self-evolved} states $\opsig_{A}(t) = \exp{\left( - i \ham_{A} t \right)} \oprho_{A}(0) \exp{\left(  i \ham_{A} t \right)}$ and $\opsig_{B}(t) = \exp{\left( - i \ham_{B} t \right)} \oprho_{B}(0) \exp{\left(  i \ham_{B} t \right)}$ and write the state $\oprho(t)$ as,
\begin{equation}
\label{rho_t_1}
\oprho(t) = \exp{\left( - i \E(t) t \right)} \left( \opsig_{A} \otimes \opsig_{B} \right) \exp{\left(  i \E(t) t \right)} \: ,
\end{equation}
which can be expanded out to $\Ottwo$ as,
\begin{equation}
\label{rho_t_2}
\oprho(t) = \left( \opsig_{A} \otimes \opsig_{B} \right)_{\Ottwo} - it \left[ \E(t) , \left( \opsig_{A} \otimes \opsig_{B} \right) \right] + \frac{(-it)^2}{2} \left[ \E(t) , \left[ \E(t), \opsig_{A} \otimes \opsig_{B} \right]\right] +  \Otthree \: .
\end{equation}
Let us now focus on one subsystem, say $\subA$, and look at its reduced dynamics by computing its reduced density matrix $\oprho_{A}(t)$ by tracing out $\subB$,
\begin{equation}
\begin{split}
\oprho_{A}(t) = \trace_{B} \oprho(t) =  \opsig_{A}(t) - it \trace_{B} \left[ \intham + \frac{i t}{2} \left[ \intham,\selfham \right] , \left( \opsig_{A} \otimes \opsig_{B} \right) \right] \\
 - \: \frac{t^2}{2} \trace_{B} \left[ \intham , \left[ \intham, \oprho(0) \right]\right] + \Otthree \: .
\end{split}
\label{rhoA_t_1}
\end{equation}
Written out (almost) explicitly using the diagonal form of $\intham$ of Eq. (\ref{Hint_diagonal}), $\oprho_{A}(t)$ takes the following form,
\begin{equation}
\label{rhoA_t_2}
\begin{split}
\oprho_{A}(t) = \opsig_{A}(t) - i t \sum_{\alpha} \lambda_{\alpha} \trace_{B}\left( \Aa \opsig_{A} \tp \B_{\alpha} \opsig_{B}  - \opsig_{A}\Aa \tp \opsig_{B}\B_{\alpha} \right) \\
 + \frac{t^2}{2}\sum_{\alpha} \lambda_{\alpha} \trace_{B} \left[ \left[ \Aa \tp \B_{\alpha} , \selfham \right] , \opsig_{A}(t) \otimes \opsig_{B}(t) \right] \\
  - \frac{t^2}{2} \sum_{\alpha,\beta} \lambda_{\alpha} \lambda_{\beta} \trace_{B} \left[ \Aa \tp \Bb , \left[ \Aa \tp \Bb , \oprho(0) \right] \right] + \Otthree \: .
\end{split}
\end{equation}
The partial trace over $\subB$ can be used to condense terms into expectation values of operators that act only on $\subB$  {for a given state $\oprho_{B}$} since $\trace_{B} \left(\hat{O}_{B} \oprho_{B}\right) = \lla \hat{O}_{B} \rra$. Let us compactly write, $\oprho_{A}(t) = \opsig_{A}(t)_{\Ottwo} + T_{1} + T_{2} + T_{3}$, which can be simplified as,
\begin{equation}
\begin{split}
T_{1} =  - i t \sum_{\alpha} \lambda_{\alpha} \trace_{B}\left( \Aa \opsig_{A} \tp \B_{\alpha} \opsig_{B}  - \opsig_{A}\Aa \tp \opsig_{B}\B_{\alpha} \right) + \Otthree \: \\
= \: -i t \sum_{\alpha} \lambda_{\alpha} \left( \lcb \Aa , \opsig_{A}(t) \rcb \lla \B_{\alpha} \rra ^{\mathrm{self}}(t)  \right)  \: ,
\label{T1}
\end{split}
\end{equation}
where $\lla \B_{\alpha} \rra ^{\mathrm{self}}_{t} = \trace_{B} \left( \B_{\alpha} \opsig_{B}(t) \right)$. We can write the other terms $T_{2}$ and $T_{3}$ to $\Ottwo$ as,
\begin{equation}
\label{T2}
T_{2} = \frac{t^2}{2} \sum_{\alpha} \lambda_{\alpha} \left( \lcb \lcb \Aa,\ham_{A} \rcb , \oprho_{A}(0)  \rcb \lla \B_{\alpha} \rra_{0}  + \lcb \Aa,\oprho_{A}(0) \rcb \lla \lcb \B_{\alpha} , \ham_{B} \rcb \rra_{0} \right) \: ,
\end{equation}
and,
\begin{equation}
\label{T3}
\begin{split}
T_{3} = \frac{-t^2}{2}  \sum_{\alpha, \beta} \lambda_{\alpha}\lambda_{\beta} \biggl(  \Aa \A_{\beta}  \oprho_{A}(0) \lla \B_{\alpha} \Bb\rra_{0} - \A_{\beta} \oprho_{A}(0) \Aa \lla \B_{\alpha} \Bb \rra_{0} - \Aa \oprho_{A}(0) \A_{\beta} \lla \Bb \B_{\alpha} \rra_{0} \\
 + \oprho_{A}(0) \A_{\beta} \Aa \lla \Bb \B_{\alpha} \rra_{0} \biggr)  \: .
\end{split}
\end{equation}
 
We next consider entanglement between the two subsystems $\subA$ and $\subB$. A common measure is to use the von-Neumann entanglement entropy $S_{vN} (\oprho) = - \Tr \left( \oprho \log{\oprho} \right)$ for a given density matrix $\oprho$. However, the presence of the logarithm makes the entropy hard to analytically compute and give expressions for, hence we will focus on its leading order contribution, the linear entropy (which is the Tsallis second order entropy measure), $S_{lin}(\oprho) = \left( 1 - \Tr \oprho^{2} \right)$. 

We can expand the self-evolved density matrix $\opsig_{A}(t)$ to $\Ottwo$ as,
 \begin{equation}
 \label{sigA_t2}
 \opsig_{A}(t) = \oprho_{A}(0) - it \lcb \ham_{A} , \oprho_{A}(0) \rcb + \frac{(-it)^2}{2} \lcb  \ham_{A}, \lcb \ham_{A} ,\oprho_{A}(0) \rcb \rcb + \Otthree \: .
 \end{equation}
 It can be explicitly checked that despite truncation upto $\Ottwo$, in each order of the expansion, the self-evolved density operator $\opsig_{A}(t)$ is pure and obeys $\opsig^{2}_{A}(t) = \opsig_{A}(t)$ and $\Tr\opsig_{A}(t) = 1$. 

 Let us now compute the linear entanglement entropy $S_{lin}(\oprho_{A}(t)) = \left( 1 - \Tr \oprho^{2}_{A}(t) \right)$ for the reduced density matrix of $\subA$ given by Eq. (\ref{rho_t_2}), which corresponds to starting with an unentangled (and hence, zero entropy) state $\oprho(0)$. Using the cyclic property of trace, it can be shown that $\Tr \left( \opsig_{A}(t) T_{1} \right) = \Tr \left( \opsig_{A}(t) T_{2} \right) = 0$ to $\Ottwo$ , and hence we get,
 \begin{equation}
 \label{Slin_1}
 S_{lin}(\oprho_{A}(t))  = 1 - \Tr \left( \opsig^{2}_{A}(t) \right) - \Tr\left( T^{2}_{1} \right) -  \Tr \left( \opsig_{A}(t) T_{3} \right)  + \Otthree \: ,
 \end{equation}
which further using $\Tr\opsig_{A}(t) = \Tr\opsig^{2}_{A}(t) = 1$ reduces to,
 \begin{equation}
 \label{Slin_2}
 S_{lin}(\oprho_{A}(t))  = - \Tr\left( T^{2}_{1} \right) -  \Tr \left( \opsig_{A}(t) T_{3} \right)  + \Otthree \: .
 \end{equation}
As we will do below -- since we are working to $\Ottwo$ -- we will replace $\opsig_{A}(t)$ with $ \oprho_{A}(0)$ in any terms that have a factor of $t^{2}$ out-front. The remaining two terms in Eq. (\ref{Slin_2}) can be computed to $\Ottwo$ in a straightforward way,
\begin{equation}
\label{TrT1sq_1}
\Tr\left( T^{2}_{1} \right)  = (-it)^{2} \sum_{\alpha,\beta} \lambda_{\alpha}\lambda_{\beta} \lla \B_{\alpha} \rra_{0} \lla \Bb \rra_{0} \Tr \left( \lcb \Aa,\oprho_{A}(0) \rcb       \lcb \A_{\beta},\oprho_{A}(0) \rcb  \right) \: ,
\end{equation}
which can be simplified by noting that for pure states $\oprho_{A}(0) = \ket{\psi_{A}(0)} \bra{\psi_{A}(0)}$, certain trace terms simplify into product of expectation values, such as,
\begin{equation}
\Tr \left( \Aa \oprho_{A}(0)  \A_{\beta} \oprho_{A}(0)  \right) =  \lla \Aa  \rra_{0} \lla \A_{\beta} \rra_{0} \: .
\end{equation}

Thus, further using such simplifications, we arrive at the following expressions for $\Tr\left( T^{2}_{1} \right)$ and $\Tr \left( \opsig_{A}(t) T_{3} \right)$ to $\Ottwo$,
\begin{equation}
\label{TrT1sq}
\Tr\left( T^{2}_{1} \right)  = -t^2 \sum_{\alpha,\beta} \lambda_{\alpha}\lambda_{\beta} \lla \B_{\alpha} \rra_{0} \lla \Bb \rra_{0} \left( 2\lla \Aa  \rra_{0} \lla \A_{\beta} \rra_{0} - \lla \{ \Aa ,\A_{\beta} \}_{+}  \rra_{0} \right)  \: ,
\end{equation}
\begin{equation}
\label{TrT3rhoA}
\begin{split}
\Tr\left( \oprho_{A}(0) T_{3} \right)  = \Tr\left( \opsig_{A}(t) T_{3} \right) =  -\frac{t^2}{2}\sum_{\alpha,\beta} \lambda_{\alpha}\lambda_{\beta}  \biggl( \lla \B_{\alpha}\Bb\rra_{0}  \lla \Aa \A_{\beta} \rra_{0}  - \lla \B_{\alpha}\Bb\rra_{0}  \lla \Aa \rra_{0} \lla\A_{\beta} \rra_{0} \\
 - \lla \Bb \B_{\alpha}\rra_{0} \lla \Aa \rra_{0} \lla\A_{\beta} \rra_{0} + \lla \Bb \B_{\alpha}\rra_{0}  \lla \Aa \A_{\beta} \rra_{0}  \biggr) \: ,
\end{split}
\end{equation}
where $\{ \hat{O}_{1},\hat{O}_{2} \}_{+} =  \left(\hat{O}_{1} \hat{O}_{2} + \hat{O}_{2} \hat{O}_{1}\right)$ is the anticommutator of $\hat{O}_{1}$ and $\hat{O}_{2}$. Putting these together in Eq. (\ref{Slin_2}), we obtain the desired result of Eq. (\ref{Slin}),
 \begin{equation}
 \label{Slin2}
 \begin{split}
 S_{lin}(\oprho_{A}(t)) \:  = \:  t^2  \sum_{\alpha,\beta} \lambda_{\alpha}\lambda_{\beta} \biggl(  \lla \Aa \A_{\beta} \rra_{0} \lla \B_{\alpha} \Bb \rra_{0} +   \lla \A_{\beta} \Aa \rra_{0} \lla  \Bb \B_{\alpha} \rra_{0}    \\
  - \lla \Aa \rra_{0} \lla \A_{\beta} \rra_{0} \left( \lla \{ \B_{\alpha} ,\Bb \}_{+} \rra_{0} -  \lla \B_{\alpha} \rra_{0} \lla \Bb \rra_{0} \right) \\
   -   \lla \B_{\alpha} \rra_{0} \lla \Bb \rra_{0} \left( \lla \{ \A_{\alpha} ,\A_{\beta} \}_{+} \rra_{0} - \lla \Aa \rra_{0} \lla \A_{\beta} \rra_{0} \right) \biggr) + \Otthree  \:  \: .
 \end{split}
 \end{equation}

\section{Generalized Pauli Operators}
\label{app:gca}

Here, we provide a brief review of generalized Pauli operators (GPOs) and their use to define finite-dimensional conjugate variables closely following the exposition of Ref. \cite{Singh:2018qzk}. The interested reader is referred to Refs. \cite{Jagannathan:1981ri,SanthanamTekumalla1976,Singh:2018qzk} (and references therein) for more detail. 

Consider a finite-dimensional Hilbert Space $\hs$ of dimension $\Dim \hs = d \in \mathbb{Z}^{+}$ with $d < \infty$. The GPO algebra on the space of linear operators $\mathcal{L}(\hs)$ acting on $\hs$ comes equipped with two unitary (but not necessarily Hermitian) operators as generators of the algebra, call them $\hat{A}$ and $\hat{B}$, which satisfy the following commutation relation,
\begin{equation}
\hat{A}\hat{B} = \omega^{-1}\hat{B}\hat{A} \: ,
\end{equation}
where $\omega = \exp\left(2 \pi i /d\right)$ is the $d$-th primitive root of unity. This commutation relation is also more commonly known as the Weyl braiding relation \cite{weyl1950theory}, and any further notions of commutations between conjugate, self-adjoint operators defined from $\hat{A}$ and $\hat{B}$ will be derived from this relation. In addition to being unitary, $\hat{A}\hat{A}^{\dag} = \hat{A}^{\dag}\hat{A} =  \eye  = \hat{B}\hat{B}^{\dag} = \hat{B}^{\dag}\hat{B}$, the algebra cyclically closes, giving it a cyclic structure in eigenspace,
\begin{equation}
\label{AdBdI}
\hat{A}^{d} = \hat{B}^d = \eye \: ,
\end{equation} 
where $\eye$ is the identity operator on $\mathcal{L}(\hs)$. 

The GPO algebra can be constructed for both even and odd values of $d$ and both cases are important and useful in different contexts. Here, we focus on the case of odd $d \equiv 2 l + 1$, which will be useful in constructing conjugate variables whose eigenvalues can be thought of labeling lattice sites, centered around zero. While the subsequent construction can be done in a basis-independent way, we choose a hybrid route, switching between an explicit representation of the GPO and abstract vector space relations. Let us follow the convention that all indices used in this section(for the case of odd $d = 2l+ 1$), for labeling states or matrix elements of an operator in some basis will run from $-l ,(-l + 1) ,\ldots,-1,0,1,\ldots,l$. The operators are further specified by their eigenvalue spectrum, and it is identical for both the GPO generators $\hat{A}$ and $\hat{B}$,
\begin{equation}
\label{specABodd}
\mathrm{spec}(\A) \: = \: \mathrm{spec}(\B) \: = \: \{\omega^{-l}, \omega^{-l + 1},\ldots, \omega^{-1}, 1, \omega^{1}, \ldots, \omega^{l-1}, \omega^{l}  \} \: .
\end{equation}

There exists a unique irreducible representation (up to unitary equivalences) (see \cite{Jagannathan:2010sb} for details) of the generators of the GPO defined via Eqs. (\ref{weylbraid}) and (\ref{AdBdI}) in terms of $N \times N$ matrices
\begin{equation}
\label{Amatrix}
  A \: = \:      \begin{bmatrix}
       0  & 0  & 0 & \ldots  & 1          \\[0.3em]
        1  & 0  & 0 &  \ldots   & 0          \\[0.3em]
        0  & 1  & 0 &  \ldots   & 0          \\[0.3em]
       . & .  & \ldots   & .          \\[0.3em]
              .  & .  & \ldots   & .          \\[0.3em]
        0  & 0   & \ldots   & 1 & 0          \\[0.3em]
     \end{bmatrix}_{N \times N} \: .
\end{equation}
\begin{equation}
\label{Bmatrix}
  B \: = \:      \begin{bmatrix}
       \omega^{-l}  & 0  & 0 & \ldots  & 0          \\[0.3em]
        0  & \omega^{-l+1}  & 0 &  \ldots   & 0          \\[0.3em]
       . & .  & \ldots   & .          \\[0.3em]
              .  & .  & \ldots   & .          \\[0.3em]
        0  & 0  & 0 & \ldots   & \omega^{l}          \\[0.3em]
     \end{bmatrix}_{N \times N} \: .
\end{equation}
The $\hat{.}$ has been removed to stress that these matrices are representations of the operators $\A$ and $\B$ in a particular basis, in this case, the eigenbasis $\B$ (so that $B$ is diagonal). More compactly, the matrix elements of operators $\A$ and $\B$ in the basis representation of eigenstates of $\B$,
\begin{equation}
\left[ A \right]_{jk} \equiv \braket{b_{j} | \A | b_{k}} =  \delta_{j,k+1} \: ,
\end{equation}
\begin{equation}
\left[ B \right]_{jk} \equiv \braket{b_{j} | \B | b_{k}}  = \omega^{j} \delta_{j,k} \: ,
\end{equation}
where is the Kronecker Delta function. 
The operator $\hat{A}$ acts as a a ``cyclic shift" operator for the eigenstates of $\B$, sending an eigenstate to the next,
\begin{equation}
\label{Aaction}
\hat{A}\ket{b_j} = \ket{b_{j+1}} \: .
\end{equation}

The unitary nature of these generators implies a cyclic structure which identifies $\ket{b_{l+1}} \equiv \ket{b_{-l}}$, so that $\hat{A}\ket{b_l} = \ket{b_{-l}}$. The operators $\A$ and $\B$ have the same relative action on the other's eigenstates, since nothing in the algebra sets the two apart. It has already been seen in Eq. (\ref{Aaction}) that $\A$ generates (unitary, cyclic) unit shifts in eigenstates of $B$ and the opposite holds too: the operator $\B$ generates unit shifts in eigenstates of $\A$ (given by the relation $\hat{A} \ket{a_k} = \omega^k \ket{a_k} \: , \: k = -l , \ldots 0, \ldots l$) and has a similar action with a cyclic correspondence to ensure unitarity,
\begin{equation}
\label{BactionOnAstates}
\hat{B}\ket{a_k} = \ket{a_{k+1}} \: ,
\end{equation}
with cyclic identification $\ket{a_{l+1}} \equiv \ket{a_{-l}}$. Hence we already have a set of operators that generate shifts in the eigenstates of the other, which is precisely what conjugate variables do and which is why we see that the GPOs provides a very natural structure to define conjugate variables on Hilbert Space.
The GPO generators $\A$ and $\B$ have been extensively studied in various contexts in quantum mechanics, and offer a higher dimensional, non-Hermitian generalization of the Pauli matrices. In particular, for $d = 2$ it will be seen that $A = \sigma_{1}$ and $B = \sigma_{3}$, which recovers the Pauli matrices. 

The defining notion for a pair of conjugate variables is the identification of two self-adjoint operators acting on Hilbert space, each of which generates translations in the eigenstates of the other. For instance, in (conventionally infinite-dimensional) textbook quantum mechanics, the momentum operator $\hat{p}$ generates shifts/translations in the eigenstates of its conjugate variable, the position $\hat{q}$ operator, and vice versa. Taking this as our defining criterion, we define a pair of Hermitian conjugate operators $\opphi$ and $\oppi$, acting on a finite-dimensional Hilbert space, each of which is the generator of translations in the eigenstates of its conjugate, with the following identification,
\begin{equation}
\A \equiv \exp{(-i \alpha \oppi)} \: , \: \: \: \: \B = \exp{(i \beta \opphi)} \: ,
\end{equation}
where $\alpha$ and $\beta$ are non-zero real parameters.

To further reinforce this conjugacy relation between operators $\A$ and $\B$, we see that they are connected to each under a discrete Fourier transformation implemented by Sylvester's Circulant Matrix $S$, which is a $N \times N$ unitary matrix $\left(SS^{\dag} = S^{\dag}S = \eye \right)$, connecting $A$ and $B$,
\begin{equation}
\label{sylvesterrelation}
S A S^{-1} = B \: .
\end{equation}
Sylvester's matrix has the following form, which we identify to be in the $\{\ket{b_j}\}$ basis, with $j$ and $k$ running from $-l, \cdots , 0, \cdots, l$:
\begin{equation}
\label{sylvestermartix}
\left[ S \right] _{jk} = \frac{\omega^{jk}}{\sqrt{N}} \: .
\end{equation}
Since $A$ and $B$ are non-singular and diagonalizable, it follows that $\log A$ and $\log B$ exist, even though multivalued. In the case of odd dimension $d = 2 l + 1$, their principle logarithms are well defined and we are able to find explicit matrix representations for operators $\opphi$ and $\oppi$. In particular, we can obtain matrix representation for $\oppi$ in the $\ket{\phi_{j}}$ basis,
\begin{equation}
\label{pimatrix}
\braket{\phi_{j} | \oppi | \phi_{j'}} = \left( \frac{ 2 \pi}{(2 l + 1)^{2} \alpha} \right) \sum_{n = -l}^{l} n \exp{\left( \frac{2 \pi i  (j - j') n }{2l + 1} \right)} \: = \:  \begin{cases}
        0 \: , &  \text{if } j = j' \\
        \\
       \left( \frac{ i \pi}{(2 l + 1) \alpha} \right) \text{cosec}{\left(\frac{2 \pi l (j - j')}{2l + 1}  \right)} \: , & \text{if } j \neq j'\;.
        \end{cases},
\end{equation}

The matrix elements of $\oppi$ in the eigenbasis of $\opphi$ are non-local, in the sense that they have power-law-like decay in $(j-j')$, and hence connect arbitrary ``far" eigenstates of $\opphi$. This is a feature of the finite-dimensional construction and in the infinite-dimensional limit $d \to \infty$, we recover the local form of $\opphi$ as $-i d/d\phi$ as expected.
 Of course, $\opphi$ has common eigenstates with those of $\B$ and $\oppi$ shares eigenstates with $\A$. The corresponding eigenvalue equations for $\opphi$ and $\oppi$ can be easily deduced using Eqs. (\ref{phipidef}) and (\ref{specABodd}),
\begin{equation}
\label{phi_eig}
\opphi \ket{\phi_j} = j \left( \frac{ 2 \pi}{(2 l + 1) \beta} \right) \ket{\phi_j} \: , \: \: \: j = -l, \ldots , 0,\ldots,l \: ,
\end{equation}
\begin{equation}
\label{pi_eig}
\oppi \ket{\pi_j} = j \left( \frac{ 2 \pi}{(2 l + 1) \alpha} \right) \ket{\pi_j} \: , \: \: \: j = -l, \ldots , 0,\ldots,l \: ,
\end{equation}

These conjugate variables defined on a finite-dimensional Hilbert space will not satisfy Heisenberg canonical commutation relation $ \left[ \opphi,\oppi \right] = i $ (in units where $\hbar = 1$), since by the Stone-von Neumann theorem there are no finite-dimensional representations of Heisenberg CCR. However, $\opphi$ and $\oppi$ still serve as a robust notion of conjugate variables and their commutation can be derived from the more fundamental Weyl Braiding Relation of Eq. (\ref{weylbraid}). In the large dimension limit $d \to \infty$, one recovers Heisenberg form of the CCR if the parameters $\alpha$ and $\beta$ are constrained to obey $\alpha \beta = 2\pi/d$.

\section{Generic Evolution of Reduced Density Operators}
\label{app:non-generic}

We can further illustrate how decoherence is a non-generic feature as discussed in Section~\ref{sec:decoherence_feature} by taking the general expression found for the reduced density operator $\oprho_{A}(t)$ to $\Ottwo$ in the bipartite case discussed in Eq. (\ref{rhoA_t_2}) and studying it further to find conditions when off-diagonal elements in the pointer basis get suppressed relatively quickly leading to effective decoherence. 

Let us compute the time derivative of the reduced density matrix, $\dot{\oprho}_{A}(t)$ to help us understand when decoherence is effective and leads to dynamic suppression of off-diagonal elements in the pointer basis. We will work explicitly to $\Ot$ to keep a tractable number of terms, enough to help us see decoherence in action,
\begin{equation}
\label{rhoAdot_1}
\rhodot(t) = \dot{\opsig}_{A}(t) + \dot{T}_{1} + \dot{T}_{2} + \dot{T}_{3} + \Ottwo \: ,
\end{equation}
where we can use the von-Neumann evolution equation for a density operator, $\dot{\opsig}_{A}(t) = - i \lcb \ham_{A} , \opsig_{A}(t) \rcb$. The time derivatives of $T_{2}$ and $T_{3}$ are easy to take from Eqs. (\ref{T2}) and (\ref{T3}) since they both have a factor of $t^2$ out-front. The time derivative of $T_{1}$ can be computed to $\Ot$ as follows,
\begin{equation}
\begin{split}
\dot{T}_{1} = \left( - i  \sum_{\alpha} \lambda_{\alpha} \lla \B_{\alpha} \rra^{\mathrm{self}}_{t} \lcb \Aa, \opsig_{A}(t) \rcb \right) - it \sum_{\alpha} \lambda_{\alpha} \lla \dot{\B}_{\alpha} \rra_{0} \lcb \Aa, \oprho_{A}(0) \rcb \\
- it \sum_{\alpha}\lambda_{\alpha} \lla \B_{\alpha} \rra_{0} \lcb \Aa,\dot{\opsig}_{A}(t) \rcb + \Ottwo \: ,
\end{split}
\label{T1dot_1}
\end{equation}
where to to retain Eq. (\ref{T1dot_1}) to $\Ot$, we can write $\dot{\opsig}_{A}(t)_{\Ot} = -i \lcb \ham_{A} , \oprho_{A}(0)\rcb$ and from the expression for $\lla \B_{\alpha} \rra^{\mathrm{self}}_{t}$,
\begin{equation}
\begin{split}
\lla \B_{\alpha} \rra^{\mathrm{self}}_{t} = \trace \left( \opsig_{B}(t) \B_{\alpha} \right) = \lla \B_{\alpha} \rra_{0} - it \trace\left( \lcb \ham_{B},\oprho_{B}(0) \rcb \B_{\alpha} \right) \\
- \frac{t^2}{2} \trace \left( \lcb \ham_{B} , \lcb \ham_{B},\oprho_{B}(0) \rcb \rcb \B_{\alpha} \right) + \Otthree \: ,
\end{split}
\label{Balpha_seft_t}
\end{equation}
we can extract the $\lla \dot{\B}_{\alpha} \rra_{0}$ which will contribute to Eq. (\ref{T1dot_1}) to $\Ot$,
\begin{equation}
\label{Bdot_alpha_0}
\lla \dot{\B}_{\alpha} \rra_{0} = i \lla \lcb \ham_{B} , \B_{\alpha} \rcb \rra_{0} \: .
\end{equation}
Plugging these in Eq. (\ref{rhoAdot_1}), we that find the term with $\lla \dot{\B}_{\alpha} \rra_{0}$ cancels with one of the terms in $\dot{T}_{2}$ to yield,
\begin{equation}
\begin{split}
\rhodot(t) = -i \lcb \ham^{\mathrm{eff}}_{A}(t) , \opsig_{A}(t) \rcb  - t  \sum_{\alpha} \lambda_{\alpha} \lla \B_{\alpha} \rra_{0} \left( \lcb \Aa, \lcb \ham_{A} , \oprho_{A}(0) \rcb \rcb - \lcb \lcb \Aa,\ham_{A}\rcb , \oprho_{A}(0)  \rcb \right) \\
+ \dot{T}_{3} + \Ottwo \: ,
\end{split}
\label{rhoAdot_2}
\end{equation}
where we have defined an  {effective} self-Hamiltonian for $\subA$, which weighs in a relevant contribution from the interaction terms $\Aa$,
\begin{equation}
\label{Heff_A}
\ham^{\mathrm{eff}}_{A}(t) = \ham_{A} + \sum_{\alpha} \lambda_{\alpha} \lla \B_{\alpha} \rra^{\mathrm{self}}_{t} \Aa  + \Ottwo \: .
\end{equation}

Let us write this in a more suggestive way such that the evolution equation of $\rhodot(t)$ can be explicitly split into a unitary piece and a piece that will induce decoherence under the right conditions. To $\Ot$, let us write $\opsig_{A}(t) = \oprho_{A}(t)_{\Ot} - \left(T_{1}\right)_{\Ot}$ and substitute in Eq. (\ref{rhoAdot_2}) while also noticing that the term $\left( \lcb \Aa, \lcb \ham_{A} , \oprho_{A}(0) \rcb \rcb - \lcb \lcb \Aa,\ham_{A}\rcb , \oprho_{A}(0)  \rcb \right)$ condenses to $\lcb \ham_{A} , \lcb \Aa , \oprho_{A}(0) \rcb \rcb$,
\begin{equation}
\label{rhoAdot_3}
\begin{split}
\rhodot(t) = -i \lcb \ham^{\mathrm{eff}}_{A}(t) , \oprho_{A}(t) \rcb  + t \sum_{\alpha} \lambda_{\alpha} \lla \B_{\alpha} \rra_{0} \lcb \ham_{A} , \lcb \Aa , \oprho_{A}(0) \rcb \rcb \\
+ t \sum_{\alpha,\beta} \lambda_{\alpha} \lambda_{\beta}  \lla \B_{\alpha} \rra_{0}  \lla \Bb \rra_{0} \lcb \Aa, \lcb \A_{\beta} , \oprho_{A}(0) \rcb \rcb \\
- t \sum_{\alpha} \lambda_{\alpha} \lla \B_{\alpha} \rra_{0} \lcb \ham_{A} , \lcb \Aa , \oprho_{A}(0) \rcb \rcb + \dot{T}_{3} + \Ottwo \: .
\end{split}
\end{equation}
The term containing $\lcb \ham_{A} , \lcb \Aa , \oprho_{A}(0) \rcb \rcb$ cancels away and after substituting for $\dot{T}_{3}$ from Eq. (\ref{T3}) and collecting terms, we see that the final expression for $\rhodot(t) $ to $\Ot$ is,
\begin{equation}
\begin{split}
\label{rhoAdot_final}
\rhodot(t) = -i \lcb \ham^{\mathrm{eff}}_{A}(t) , \oprho_{A}(t) \rcb   - t \sum_{\alpha,\beta} \lambda_{\alpha} \lambda_{\beta} \biggl( \left(\Aa\A_{\beta} \oprho_{A}(0) - \A_{\beta} \oprho_{A}(0)\Aa \right) \left( \lla \B_{\alpha}\Bb \rra_{0} - \lla \B_{\alpha}\rra_{0}\lla \Bb \rra_{0} \right)      \\
  +      \left( \oprho_{A}(0) \A_{\beta} \Aa - \Aa \oprho_{A}(0)\A_{\beta} \right) \left( \lla \Bb\B_{\alpha} \rra_{0} - \lla \Bb\rra_{0}\lla \B_{\alpha} \rra_{0} \right)   \biggr) + \Ottwo \: .
\end{split}
\end{equation}
Thus, we see that the equation for $\rhodot(t)$ to $\Ot$ splits into a term $ \left(-i \lcb \ham^{\mathrm{eff}}_{A}(t) , \oprho_{A}(t) \rcb \right)$, which corresponds to  {unitary} evolution of $\oprho_{A}(t)$ under the effective self-Hamiltonian $\ham^{\mathrm{eff}}_{A}(t)$ and a term that will be responsible for decoherence under right conditions. 

Let us focus on this ``decoherence" term $\mathcal{D}(\oprho_{A})$ and not concern ourselves with the unitary evolution for the moment (the $\supset$ representing that we are focusing only on the decoherence term),
\begin{equation}
\label{decoherence_term}
\begin{split}
\rhodot(t) \supset \mathcal{D}(\oprho_{A}) + \Ottwo \equiv  - t \sum_{\alpha,\beta} \lambda_{\alpha} \lambda_{\beta} \biggl( \left(\Aa\A_{\beta} \oprho_{A}(0) - \A_{\beta} \oprho_{A}(0)\Aa \right) \left( \lla \B_{\alpha}\Bb \rra_{0} - \lla \B_{\alpha}\rra_{0}\lla \Bb \rra_{0} \right)      \\
  +      \left( \oprho_{A}(0) \A_{\beta} \Aa - \Aa \oprho_{A}(0)\A_{\beta} \right) \left( \lla \Bb\B_{\alpha} \rra_{0} - \lla \Bb\rra_{0}\lla \B_{\alpha} \rra_{0} \right)   \biggr) + \Ottwo \: . \: 
  \end{split}
\end{equation}
In the Quantum Measurement Limit, when there exists a consistent {pointer basis} $\{ \ket{a_{j}} \: | \: j = 1,2,\cdots,d_{A}\}$ which will be selected such that it forms simultaneous eigenstates of  {all} $\Aa \: \: \forall \: \: \alpha$,
\begin{equation}
\label{pointer_states_eigen}
\Aa  \ket{a_{j}} \: = \: a^{\alpha}_{j} \ket{a_{j}} \: \: \forall \: \alpha \: \: \mathrm{and} \: j = 1,2,\cdots,d_{A} \: .
\end{equation} 
This is a highly non-generic situation, since an arbitrary Hamiltonian in an arbitrary factorization will have non-commuting terms in $\intham$ and hence not admit a complete basis satisfying Eq. (\ref{pointer_states_eigen}) to serve as a pointer basis. For decoherence to be effective, there would be a small number of consistent terms in $\intham$ being monitored by the other subsystem as discussed in Section \ref{subsec:min_entropy}. 

Let us see this explicitly by considering the off-diagonal matrix element $\braket{a_{j} | \rhodot(t) | a_{k} }, \: \:  j \neq k$ of $\rhodot(t)$ in the purported pointer basis $\{ \ket{a_{j}} \}  $. The decoherence term $\mathcal{D}(\oprho_{A}(t))$ in Eq. (\ref{decoherence_term}) can be further split into $\alpha = \beta$ terms and $\alpha \neq \beta$ ones. The cross-terms with $\alpha \neq \beta$ are not seen to have a definitive sign that is needed for decoherence to take place. On the other hand, let us look at the $\alpha = \beta$ terms of the matrix element,
\begin{equation}
\label{like_matrix_element_1}
\braket{a_{j} | \rhodot(t) | a_{k} } \supset - t \sum_{\alpha} \lambda^{2}_{\alpha} \left( \lla \B^{2}_{\alpha}\rra_{0} - \lla \B_{\alpha} \rra^{2}_{0} \right)  \braket{a_{j} | \left(  \Aa^{2}\oprho_{A}(0) - 2 \Aa \oprho_{A}(0)\Aa + \oprho_{A}(0) \Aa^2 \right)  | a_{k}} \: \: , j \neq k \: ,
\end{equation}
which can be further simplified using Eq. (\ref{pointer_states_eigen}),
\begin{equation}
\label{like_matrix_element_2}
\left[ \frac{d}{dt } \oprho_{A}(t)\right]_{jk} \supset - t \sum_{\alpha} \lambda^{2}_{\alpha} \left( \lla \B^{2}_{\alpha}\rra_{0} - \lla \B_{\alpha} \rra^{2}_{0} \right) \left( a_j - a_k \right)^2 \left[ \oprho_{A}(0) \right]_{jk} + \Ottwo \: .
\end{equation}
Now since we are working to $\Ot$ in Eq. (\ref{like_matrix_element_2}), we can replace $\left[ \oprho_{A}(0) \right]_{jk}$ with  $\left[ \oprho_{A}(t) \right]_{jk}$ since corrections will contribute to $\Ottwo$ due to the presence of the factor of $t$ in the expansion,
\begin{equation}
\label{like_matrix_element_3}
\left[ \frac{d}{dt } \oprho_{A}(t)\right]_{jk} \supset - t \left( \sum_{\alpha} \lambda^{2}_{\alpha} \: \Delta^{2}\left(\B_{\alpha}\right)_{0} \left( a_j - a_k \right)^2 \right) \left[ \oprho_{A}(t) \right]_{jk} + \Ottwo \: .
\end{equation}
The term in the parenthesis $\left( \sum_{\alpha} \lambda^{2}_{\alpha} \: \Delta^{2}\left(\B_{\alpha}\right)_{0} \left( a_j - a_k \right)^2 \right)$ is  {positive definite} since the term, $\Delta^{2}\left(\B_{\alpha}\right)_{0} \equiv \left( \lla \B^{2}_{\alpha}\rra_{0} - \lla \B_{\alpha} \rra^{2}_{0} \right) $ is the variance of $\B_{\alpha}$ in the state $\oprho_{A}(0)$, and hence positive by construction. This leads to decoherence since off-diagonal terms in Eq. (\ref{like_matrix_element_3}) get suppressed dynamically in the pointer basis selected by $\intham$. 

Thus, we see that for decoherence to be effective, there should exist a small number of consistent terms in $\intham$ being monitored by the other subsystems ($\subB$ in this case), which will give us a notion of pointer basis in which off-diagonal elements of $\oprho_{A}(t)$ are dynamically suppressed due to interaction with the environment. Most of our classic models of decoherence \cite{2007dqct.book.....S} indeed consist of a single term (or a small number of compatible terms) representing environmental monitoring of the form $\intham = \lambda \left(\A \otimes \B \right)$ and hence there will be decoherence in the eigenbasis of $\A$, which serve as pointer states. From Eq. (\ref{like_matrix_element_3}), we can give an estimate for the decoherence time-scale $\tau_{d}$ for the $(j,k)$ matrix element, focusing on the $\intham = \lambda \left(\A \otimes \B \right)$ for clarity,
\begin{equation}
\label{decoherence_time_1}
\left(\tau_{d} \right)_{jk} \sim  \frac{\sqrt{2}}{|\lambda| \: \: |a_j - a_k| \: \:  \left| \Delta \left( \B_{\alpha} \right)_{0} \right|}  \: .
\end{equation}
Thus, as we can see from the above Eq. (\ref{decoherence_time_1}), for higher interaction strength, there is more stronger monitoring of $\subA$ by $\subB$ and hence faster decoherence. More variance of $\B$ in the initial state allows for more support in state space for monitoring and quicker suppression of interference and also, we see that decoherence time-scales are inversely proportional to the spectral differences in $\A$. This can also be easily understood since more spacing between eigenvalues of $\A$ would lead to inducing faster orthogonality in conditional states of $\subB$, and hence more effective decoherence.
\section{Pointer Entropy Growth in the Quasi-Classical Factorization}
\label{app:pointer_ent_QML}
In this appendix, we compute $\ddot{S}_{pointer}(0)$ for the QC factorization of a given Hamiltonian (which satisfies the QML). Let us first compute $\dot{S}_{pointer}$ explicitly to help us get to $\ddot{S}_{pointer}(0)$. Since we want an expression for $\ddot{S}_{pointer}(0)$, we will just retain $\Ot$ in the following $\dot{S}_{pointer}$ calculation. From the definition of $S_{pointer}$ of Eq. (\ref{Spointer_define}), we see,
\begin{equation}
\label{Spointer_dot1}
\dot{S}_{pointer}(t) = -2 \sum_{j=1}^{d_{A}} p_{j}(t)\dot{p}_{j}(t) \: ,
\end{equation}
where $p_{j}(t)$ is the probability distribution defined by,
\begin{equation}
p_{j}(t) = \Tr_{A} \biggl( \oprho_{A}(t) \ket{a_j}\bra{a_j} \biggr) \equiv  \Tr_{A} \biggl( \oprho_{A}(t) \hat{O}_{j} \biggr) = \braket{a_{j} | \oprho_{A}(t) | a_{j} }\: ,
\end{equation}
where $\{\ket{a_{j}}\}$ is the set of eigenstates of the pointer observable $\Oa$, and $\hat{O}_{j} \equiv \ket{a_{j}}\bra{a_j}$. Following the construction in Appendix (\ref{sec:low_entropy}), we can write $\oprho_{A}(t)$ to $\Ot$ as,
\begin{equation}
\label{oprhoA_Ot}
\begin{split}
\oprho_{A}(t) & = \oprho_{A}(0) - it \lcb \ham_{A} , \oprho_{A}(0) \rcb  - it \sum_{\alpha}\lambda_{\alpha} \lla\B_{\alpha}\rra_{0} \lcb \Aa,\oprho_{A}(0) \rcb \\
& \equiv \oprho_{A}(0) - it \lcb \ham^{\mathrm{eff}}_{A}(0),\oprho_{A}(0) \rcb + \Ottwo \: ,
\end{split}
\end{equation}
from which we get,
\begin{equation}
\label{pj_1_full}
p_{j}(t) = \Tr\left( \oprho_{A}(t) \Oj \right) = p_{j}(0) - it\lla \lcb \Oj , \ham^{\mathrm{eff}}_{A}(0)\rcb \rra_{0} + \Ottwo \: .
\end{equation}
To $\Ot$ in the above equation, the effective self-Hamiltonian $\ham^{\mathrm{eff}}_{A}(0)$ contains a contribution from the interaction terms,
\begin{equation}
\ham^{\mathrm{eff}}_{A}(0) = \ham_{A} + \sum_{\alpha} \lambda_{\alpha} \lla \B_{\alpha} \rra_{0} \Aa   \: .
\end{equation}
The pointer observable in the QML satisfies $\lcb \Oj,\Aa \rcb = 0 \: \:\forall \: \: \alpha, j$, this can be simplified further to depend only on $\ham_{A}$,
\begin{equation}
\label{pj_1_qml}
p_{j}(t) = p_{j}(0) - it\lla \lcb \Oj , \ham_{A}\rcb \rra_{0} + \Ottwo  \: \: \: \: \: \:  \mathrm{(QML)}\: ,
\end{equation}
where $p_{j}(0) = \lla\Oj\rra_{0}$. 

To compute $\dot{S}_{pointer}(t)$, we use Eq. (\ref{rhoAdot_final}) for $d\oprho_{A}/dt$ to $\Ot$ and notice that, as remarked in section \ref{sec:pointer}, the diagonal entries of the decoherence term $\mathcal{D}(\oprho_{A}(t))$ in the pointer basis vanish identically, giving the diagonal entries of $d\oprho_{A}/dt$ in the pointer basis as shown in Eq. (\ref{rhodot_diagonal}),
\begin{equation}
\left[ \frac{d}{dt } \oprho_{A}(t)\right]_{jj} = \biggl(-i \lcb \ham_{A}(t) , \oprho_{A}(t) \rcb_{jj}\biggr) + \Ottwo \: ,
\end{equation}
which gives us,
\begin{equation}
\label{pjdot_1_full}
\begin{split}
\dot{p}_{j}(t) & = \Tr\left(\Oj \: \frac{d}{dt } \oprho_{A}(t)  \right) \\
&= -i \Tr\left(\lcb\ham_{A},\oprho_{A}(t) \rcb \Oj \right) + \Ottwo \: .
\end{split}
\end{equation}
Substituting for $\oprho_{A}(t)$ to $\Ot$ from Eq. (\ref{oprhoA_Ot}), we get,
\begin{equation}
\label{pjdot_2_full}
\dot{p}_{j}(t) = \dot{p}_{j}(0) - t\Tr\left(\lcb \ham_{A},\lcb \ham_{A},\oprho_{A}(0)\rcb\rcb \Oj\right) - t\sum_{\alpha}\lambda_{\alpha}\lla\B_{\alpha}\rra_{0} \Tr\left( \lcb \ham_{A},\lcb \Aa,\oprho_{A}(0)\rcb\rcb \Oj\right) + \Ottwo \: ,
\end{equation}
where $\dot{p}_{j}(0) = -i \Tr\left( \lcb \ham_{A},\oprho_{A}(0)\rcb \Oj\right) = -i \lla \lcb \Oj,\ham_{A}\rcb \rra_{0} $. We can now further simplify this in the Quantum Measurement Limit when a consistent pointer observable exists, and after a few lines of trace manipulations we obtain,
\begin{equation}
\label{pjdot_1_QML}
\dot{p}_{j}(t) = \dot{p}_{j}(0) - t \lla \Oj \ham^{2}_{A} + \ham^{2}_{A}\Oj - 2\ham_{A}\Oj\ham_{A}\rra_{0} - t\sum_{\alpha}\lambda_{\alpha} \lla \B_{\alpha} \rra_{0} \lla\lcb\Oj,\lcb\ham_{A},\Aa\rcb\rcb \rra_{0} + \Ottwo \: .
\end{equation}
We can now string everything together to give us an expression for $\ddot{S}_{pointer}(0)$ for the pointer observable $\Oa$, by taking a time derivative of $\dot{S}_{pointer}(t)$ constructed out of Eqs. (\ref{pj_1_qml}) and (\ref{pjdot_1_QML}),
\begin{equation}
\begin{split}
\ddot{S}_{pointer}(0) = 2\sum_{j=1}^{d_{A}} \lla \lcb \Oj,\ham_{A}\rcb \rra^{2}_{0} +
2\sum_{j=1}^{d_{A}}\left( p_{j}(0) \lla\Oj \ham^{2}_{A} + \ham^{2}_{A}\Oj - 2\ham_{A}\Oj\ham_{A}\rra_{0} \right) + \\
2\sum_{j=1}^{d_{A}}\left( p_{j}(0) \sum_{\alpha}\lambda_{\alpha} \lla \B_{\alpha} \rra_{0} \lla\lcb\Oj,\lcb\ham_{A},\Aa\rcb\rcb \rra_{0}  \right)\:  .
\end{split}
\label{Spointerdoubledot_QML}
\end{equation}
\bibliographystyle{utphys}
\bibliography{Quantum_Mereology}
\end{document}